%% file: main.tex
\PassOptionsToPackage{dvipsnames}{xcolor}

\documentclass[sigconf]{acmart}

\usepackage{tikz}
\usepackage{amsmath}
\usepackage{listings}
\usepackage{xurl}
\usepackage{booktabs}
\usepackage{multirow}
\usepackage{caption}
\usepackage{subcaption}
\usepackage{graphicx}
\usepackage{tabularray}

\newif\ifdraft
\drafttrue

\input{includes/macros}

\input{includes/data}

\input{figs}
\copyrightyear{2025}
\acmYear{2025}
\setcopyright{cc}
\setcctype{by-nc}
\acmConference[CCS '25]{Proceedings of the 2025 ACM SIGSAC Conference on Computer and Communications Security}{October 13--17, 2025}{Taipei, Taiwan}
\acmBooktitle{Proceedings of the 2025 ACM SIGSAC Conference on Computer and Communications Security (CCS '25), October 13--17, 2025, Taipei, Taiwan}\acmDOI{10.1145/3719027.3744843}
\acmISBN{979-8-4007-1525-9/2025/10}
\begin{document}
%-------------------------------------------------------------------------------

\title[Local Frames: Exploiting Inherited Origins to Bypass Content Blockers]{Local Frames: Exploiting Inherited Origins\\ to Bypass Content Blockers}
% \author{USENIX Security 2025 Cycle 1 Submission \#266}

\author{Alisha Ukani}
\affiliation{%
  \institution{University of California, San Diego}
  \city{San Diego}
  \state{CA}
  \country{USA}
  }
\email{aukani@ucsd.edu}

\author{Hamed Haddadi}
\affiliation{%
  \institution{Imperial College London \& Brave Software Inc}
  \city{London}
  \country{UK}
  }
\email{h.haddadi@imperial.ac.uk}

\author{Alex C. Snoeren}
\affiliation{%
  \institution{University of California, San Diego}
  \city{San Diego}
  \state{CA}
  \country{USA}
  }
\email{snoeren@cs.ucsd.edu}

\author{Peter Snyder}
\affiliation{%
  \institution{Brave Software Inc}
  \city{San Francisco}
  \state{CA}
  \country{USA}
  }
\email{pes@brave.com}


% \ccsdesc[500]{Security and privacy~Web application security}


\input{content/abstract}
\maketitle

\input{content/intro}
\input{content/background}
\input{content/crawl_analysis}

\input{content/vuln_tools}
\input{content/ethics_disclosure}
\input{content/discussion}
\input{content/conclusion}
\input{content/acks}

% % Only show scratch pages for our drafts
% \inDraft{
%     % \input{scratch/disclosure_status}
%     \input{scratch/maximum-test-space}
% }

\bibliographystyle{plain}
\bibliography{paper}

% \appendices
% \clearpage
\appendix
\input{appendix/fp_apis}
\input{appendix/measurement}
%\input{appendix/tool_popularity}
\input{appendix/testing_details}

\end{document}

%% file: includes/macros.tex
\ifdraft
\newcommand{\todo}[1]{\textcolor{red}{\noindent[TODO: #1]}}
\else
\newcommand{\todo}[1]{}
\fi

\ifdraft
\newcommand{\alisha}[1]{\textcolor{blue}{\noindent[Alisha: #1]}}
\else
\newcommand{\alisha}[1]{}
\fi

\ifdraft
\newcommand{\alex}[1]{\textcolor{Purple}{\noindent[Alex: #1]}}
\else
\newcommand{\alex}[1]{}
\fi

\ifdraft
\newcommand{\pete}[1]{\textcolor{Green}{\noindent[Pete: #1]}}
\else
\newcommand{\pete}[1]{}
\fi

\definecolor{c1}{RGB}{254,197,82}
\definecolor{c2}{RGB}{207,79,3}

\newcommand*\emptycirc[1][1ex]{\tikz\draw (0,0) circle (#1);} 
\newcommand*\halfcirc[1][1ex]{%
	\begin{tikzpicture}
		\draw[fill] (0,0)-- (90:#1) arc (90:270:#1) -- cycle ;
		\draw (0,0) circle (#1);
\end{tikzpicture}}
\newcommand*\fullcirc[1][1ex]{\tikz\fill (0,0) circle (#1);} 

\definecolor{codegreen}{rgb}{0,0.6,0}
\definecolor{codegray}{rgb}{0.5,0.5,0.5}
\definecolor{codepurple}{rgb}{0.58,0,0.82}
\definecolor{backcolour}{rgb}{0.95,0.95,0.92}

\lstdefinestyle{mystyle}{
    backgroundcolor=\color{backcolour},   
    commentstyle=\color{codegreen},
    keywordstyle=\color{magenta},
    numberstyle=\tiny\color{codegray},
    stringstyle=\color{codepurple},
    basicstyle=\ttfamily\footnotesize,
    breakatwhitespace=false,         
    breaklines=true,                 
    captionpos=b,                    
    keepspaces=true,                 
    numbers=left,                    
    numbersep=5pt,                  
    showspaces=false,                
    showstringspaces=false,
    showtabs=false,                  
    tabsize=2,
}

\lstdefinestyle{filterrules}{
    backgroundcolor=\color{white},
    basicstyle=\ttfamily\small,
    numbers=left,
    tabsize=2,
    breakindent=10pt,
}

\newcommand{\LF}{local frame}
\newcommand{\LFs}{local frames}
\newcommand{\EG}{e.g.,}
\newcommand{\IE}{i.e.,}
\newcommand{\JS}{JavaScript}

%% file: includes/data.tex
\newcommand{\CrawlNumSitesCompletedPopular}{11,965} % updated
\newcommand{\CrawlNumSitesAttemptedMedium}{6,362} % updated
\newcommand{\CrawlNumSitesAttemptedUnpopular}{7,100} % updated
\newcommand{\CrawlNumSitesCompleted}{21,965} % updated
\newcommand{\CrawlNumSitesUncompletedPct}{23\%} % updated
\newcommand{\CrawlNumFramesAllSites}{74,263} % updated
\newcommand{\CrawlSitesWithFramesPct}{56\%} % updated
\newcommand{\CrawlSitesWithFramesNum}{12,234} % updated
\newcommand{\CrawlSitesWithFramesMean}{3.4} % updated

\newcommand{\CrawlThirdPartyFramesTotal}{28,812} % updated
\newcommand{\CrawlSitesWithThirdPartyFramesTotal}{4,776} % updated

\newcommand{\CrawlRequestsAll}{2,418,425} % updated
\newcommand{\CrawlRequestsTotal}{96,262} % updated
\newcommand{\CrawlRequestsTotalPct}{4.0\%} % updated
\newcommand{\CrawlRequestsLocalLocalFrameBlocked}{70,938} % updated
\newcommand{\CrawlRequestsLocalLocalFrameBlockedPct}{73.7\%} % updated
\newcommand{\CrawlRequestsSitesAnyRequest}{20,796} % updated
\newcommand{\CrawlRequestsSitesAnyRequestPct}{94.7\%} % updated
\newcommand{\CrawlRequestsSites}{5,168} % updated; number of sites in local frames
\newcommand{\CrawlRequestsSitesPct}{24.4\%} % updated
\newcommand{\CrawlRequestsSitesLocalFrameBlocked}{3,142} % updated
\newcommand{\CrawlRequestsSitesLocalFrameBlockedPct}{61.9\%} % updated
\newcommand{\CrawlRequestsSitesBlockedOfSites}{14.3\%} % updated
\newcommand{\CrawlRequestsSitesLocalFrameBlockedMedium}{778} % updated
\newcommand{\CrawlRequestsSitesLocalFrameBlockedUnpopular}{477} % updated
\newcommand{\CrawlRequestsSitesLocalFrameBlockedMediumPct}{59.6\%} % updated
\newcommand{\CrawlRequestsSitesLocalFrameBlockedUnpopularPct}{50.8\%} % updated

\newcommand{\CrawlFPAPIsSites}{4,280} % updated; total number of sites with local frames calling FP APIs

\newcommand{\CrawlJSCallsTotal}{111,231,650} % updated

\newcommand{\CrawlJSCallsSites}{6,596} % updated
\newcommand{\CrawlJSCallsSuspectSitesNum}{630}
\newcommand{\CrawlDwellTime}{30 seconds}

%% file: content/abstract.tex
\begin{abstract}
%Many Internet users employ content blocking tools to create a more pleasant and private Web %browsing experience.  Content blockers attempt to  hide ads and defeat user tracking and %fingerprinting through a variety of techniques.  
%We identify a class of Web functionality that is mishandled by a wide range of popular and commonly used Web privacy and security tools.  Specifically, we show that several popular content blockers are easily evaded by websites that wrap content in a \emph{local frame}: an iFrame with a source of ``about:blank''. This vulnerability allows websites to include tracking code that would otherwise be blocked and show ads that the user prefers to be hidden.

We present a study of how local frames (i.e., iframes loading content like ``about:blank'') are mishandled by a wide range of popular Web security and privacy tools. As a result, users of these tools remain vulnerable to the very attack techniques against which they seek to protect themselves, including browser fingerprinting, cookie-based tracking, and data exfiltration. The tools we study are vulnerable in different ways, but all share a root cause: legacy Web functionality interacts with browser privacy boundaries in unexpected ways, leading to systemic vulnerabilities in tools developed, maintained, and
recommended by privacy experts and activists.

%We find that AdBlock Plus, AdGuard, uBlock Origin Lite, the Brave Browser, and Safari Content Blocking all have capabilities that can be evaded by local frames. Most egregiously, we show that all of the filter list-based protections provided by Brave's iOS browser can be evaded, and that AdGuard incorrectly treats third-party local frames as first-party, allowing third parties to inject tracking scripts across all of AdGuard's supported platforms. 

We consider four core capabilities supported by most privacy tools and develop tests to determine whether each can be evaded through the use of local frames.  We apply our tests to six popular Web privacy and security tools---identifying at least one vulnerability in each for a total of 19---and extract common patterns regarding their mishandling of local frames.  Our measurement of popular websites finds that \CrawlSitesWithFramesPct{} employ local frames and that \CrawlRequestsLocalLocalFrameBlockedPct{} of the requests made by these local frames should be blocked by popular filter lists but instead trigger the vulnerabilities we identify. From another perspective, \CrawlRequestsSitesBlockedOfSites{} of all sites that we crawl make requests that should be blocked inside of local frames.  
%
%Specifically, this work makes the following contributions to the area of Web privacy and security research: i. identifying common patterns in how popular Web privacy and security tools systematically mishandle iframes with local URLs (i.e., ``local frames''), ii. defining XXX different attacks that can be conducted by exploiting local frames, iii. identifying which popular Web privacy and security tools are vulnerable to which attack methods, iv. a measurement of how commonly local frames are used by popular (i.e., Tranco 15k) websites, and v. a discussion of the 
%
%Users visiting these sites with vulnerable tools, however, are subjected to unwanted content and tracking.  
We disclosed these vulnerabilities to the tool authors
%organizations; as of this writing AdGuard, Apple, and Brave are implementing fixes.
%We practice responsible disclosure in sharing our findings with the authors of these tools,
and discuss both our experiences working with them to patch their products
%We also commit to sharing our full dataset and source code used in this research.
and the implications of our findings for other privacy and security research.

\end{abstract}

%% file: content/intro.tex
\section{Introduction}

Content-blocking tools are used by millions of people in order to protect their privacy by blocking tracking scripts, stay safe from scammers by blocking malware, save money by using less data, and enjoy a more pleasant browser experience by hiding ads. In order to provide these benefits, most content blockers maintain filter lists, which are lists of URLs that determine which network requests should be blocked or allowed.
Modern tools also implement more sophisticated defenses such as resource replacement (e.g., loading benign scripts instead of privacy-invasive ones), scriptlet injection (i.e., employing custom JavaScript code to remove cookies or block trackers), and cosmetic filtering to hide undesirable page elements.

Yet, we find that several popular content blockers are vulnerable to evasion of one or more of these capabilities---and the majority of popular websites are currently doing such evasion.
Specifically, our work shows that content blockers frequently mishandle a class of iframes we call \emph{local frames} (an iframe with a non-URL source like ``about:blank''), allowing content loaded within local frames to bypass blocker protections. Local frames initially load an empty HTML document, but content can be added to these frames dynamically. Local frames are popular with Web developers for a number of reasons, chief among them the fact that a local frame creates a clean JavaScript environment that can still access the main page. We find that content blockers fail to properly implement protections in local frames, 
%interpret both HTML requests and scripts executed in local frames. This miscalculation 
allowing websites to include tracking scripts that should be blocked and show ads that should be hidden---all by wrapping their existing code in a local frame.

Content blockers use a frame's origin---the combination of a URL's protocol (e.g. ``https''), hostname, and port---to determine how to handle its content.
For example, many content blockers provide the option to block network requests made in third-party contexts; in order to determine if a request is made in a first- or third-party context, the content blocker must know the origin of the request's source and destination.
%\todo{Make it clear that also we have an issue if rules don't get injected in local frames at all}
%
While some tools (e.g., Brave on iOS) simply fail to provide any protection in local frames, we find the most common reason why local frames evade content blockers is that content blockers mis-attribute their origin and, thus, do not associate local frames with the site that creates them.  The origin of an iframe is usually extracted from the iframe's source, but according to the HTML specification, local frames are supposed to inherit the origin of the document that creates them~\cite{mdn_origin}. For example, in Listing~\ref{lst:localframes}, the local frame on line 2 should inherit the origin of the main page (\texttt{https://firstparty.com} in this example), whereas the local frame on line 7 should inherit the origin \texttt{https://thirdparty.com}.  However, some content blockers do not correctly determine the origin of local frames; some compute the origin of the third-party local frame as ``about:blank'' while others may even fall back to \texttt{https://firstparty.com}.

\begin{lstlisting}[language=HTML, label=lst:localframes, float, floatplacement=t, deletekeywords={for, FRAME}, caption=HTML code at \texttt{https://firstparty.com} which creates two iframes: a local frame and a third-party iframe that embeds its own local frame.\vspace{-2em}]
<body>
  <iframe src="about:blank">
    <!-- Local frame for firstparty.com. Origin should be https://firstparty.com -->
  </iframe>

  <iframe src="https://thirdparty.com">
    <iframe src="about:blank">
      <!-- Local frame for thirdparty.com. Origin    should be https://thirdparty.com -->
    </iframe>
  </iframe>
</body>
\end{lstlisting}

% In particular, we consider the corner case of iFrames with a \texttt{src} attribute of \texttt{about:blank}; we call these types of iFrames \textbf{local frames}. Local frames initially load an empty HTML document, but content can be added to these frames dynamically. An advantage of creating a local frame is that it can create a clean JavaScript environment so that its content does not affect the main page and vice versa. As a result, local frames are often used for serving ads and for browser fingerprinting scripts.

%We find that 
Local frames are highly prevalent on the Web, and evasion is currently taking place (although we cannot determine if this evasion is intentional). We conduct a measurement study of websites with varying degrees of popularity according to Tranco and find local frames on \CrawlSitesWithFramesPct{} of websites (\CrawlSitesWithFramesNum{} of the \CrawlNumSitesCompleted{} successfully crawled websites). \CrawlRequestsLocalLocalFrameBlockedPct{} of the URLs requested by these local frames should be blocked by popular content blockers according to a combination of filter lists from EasyList~\cite{easylist}, EasyPrivacy~\cite{easyprivacy}, and uBlock Origin~\cite{ublockorigin}. Users browsing these sites with vulnerable tools, however, will be exposed to the content anyway.  Said another way, \CrawlRequestsSitesBlockedOfSites{} of the websites we study are evading content blockers by making requests that would otherwise be blocked, but succeed because they are made inside of local frames.  We emphasize, however, that we do not know if publishers create the local frames with an explicit intent of evading content blockers.
% 26\% of these blocked requests are made to \texttt{googlesyndication.com}, indicating that local frames are often used for ads.

% Clarify difference between origin and src
% Source doesn't always contain the origin, need to look at parent frame
% Per the HTML specification, local frames are supposed to inherit the origin of the document that creates them, but in practice, content blockers do not follow this. An origin is the combination of protocol (e.g. ``https''), hostname, and port of a URL; but having an origin of ``about:blank'' does not communicate any useful information. For example, in Listing~\ref{lst:localframes}, the local frame on line 2 should inherit the origin \texttt{https://firstparty.com}, whereas the local frame on line 7 should inherit the origin \texttt{https://thirdparty.com}.

% A content blocker cannot block the content in a local frame by blocking the \texttt{about:blank} URL, because this would also affect the local frames created by other websites. Instead, content blockers must consider the origin (i.e. the combination of scheme, hostname, and port) of the document that created the local frame. For example, in Listing~\ref{lst:localframes} we give an example of the code for \texttt{firstparty.com}, which has a local frame and also embeds an iFrame for \texttt{thirdparty.com}, which itself has a local frame. We will use this structure throughout this paper.

% For each of these capabilities, we design tests to see if local frames can trigger unintended behavior, and test a variety of popular content blocking tools.

We consider four distinct capabilities of popular blockers---request blocking, resource replacement, 
scriptlet injection, and cosmetic filtering---and
find vulnerabilities in AdBlock Plus, AdGuard, uBlock Origin Lite, the Brave Browser, DuckDuckGo, and Safari Content Blocking (which is used by many iOS apps):

\begin{itemize}
\item We find that websites can employ local frames to completely bypass filter list-based protections in Brave's iOS browser.
\item For AdGuard, we find that local frames within third-party iframes inherit the origin of the first-party website.
%; using the example in Listing~\ref{lst:localframes}, this means the local frame on line 7 would have the origin \texttt{https://firstparty.com} instead of \texttt{https://thirdparty.com}. 
This means that AdGuard users can not only be tracked by third parties, but they may also experience site breakage because rules are applied improperly in local frames. 

%The vulnerabilities we find show that there is a widespread mishandling and misunderstanding of local frames in security/privacy tools. Despite local frames appearing on 52\% of websites, we find they can be easily exploited to bypass some of the most basic capabilities of popular content blocking tools. 

%This paper makes the following contributions:
%    \item We conduct a large-scale measurement of the prevalence and behavior of local frames for the Tranco top 15k.
    
    \item Scriptlet injection and cosmetic filtering implemented in Brave's browser, as well as AdGuard's browser extensions and iOS app, can be evaded, allowing websites to perform sophisticated tracking and show ads that should be hidden.
    \item We find that the cosmetic filtering in uBlock Origin Lite, the AdBlock Plus iOS app, and Safari Content Blocking can be evaded, allowing websites to show ads that should be hidden.
    \item We find that while DuckDuckGo properly blocks tracking requests made inside local frames, they do not report these actions to the user like they do for regular frames.
\end{itemize}

We disclosed each of the 19 vulnerabilities we found to the relevant organization. Brave, Safari, AdGuard, and DuckDuckGo have patched their tools, and our disclosure has been acknowledged by uBlock Origin Lite and AdBlock Plus.
Our code and data is available at \texttt{https://osf.io/9yq57}.

%% file: content/background.tex
\section{Background and Related Work}

In this section we provide a brief overview of local frames, their intended use cases, and their relevance to content blocking.  We then introduce the four common content-blocking capabilities we study before summarizing related work.

\subsection{Local Frames}

We use the term \emph{local frame} to refer to an iframe with a non-URL source, the most common of which is ``about:blank''. This specific URI is often the default URI for new iframes~\cite{mdn_origin} and is intended to be used by browsers for blank pages~\cite{rfc6694}; the resulting local frames are uninitialized iframes.
Local frames (like all iframes) have their own page document~\cite{html}, which creates a new environment that is unaffected by the JavaScript or CSS rules scoped to the document of the main page. However, as we discuss in the next section, local frames are distinct from other iframes because they can still access the parent document (i.e., they are not fully isolated).
% Local frames are often used because they code to execute in a new environment that is unaffected by the JavaScript or CSS rules in the main page. Local frames, as with all iFrames, have their own page document~\cite{html}. This allows the contents of the local frame to operate in a new, clean environment; for example, CSS rules are local to documents, so the CSS rules in the parent frame will not automatically apply to the newly created iFrame~\cite{css}. It is important to note, however, that if frames have the same origin, they can access each other's page documents (i.e. they are not fully isolated).

\subsubsection{Inherited origins}\label{sec:bkgd:origins}

An origin for a URL is defined as the URL's protocol, port, and hostname (excluding subdomains)~\cite{mdn_origin}. Origins are used to isolate websites from potentially malicious embedded content through the \emph{same-origin policy} feature. This feature prevents a website from accessing the content of other tabs in the user's browser, or reading the cookies set for other websites (which could then be used for a cookie-hijacking attack).

However, the URL-based definition of origins cannot be applied to other URI schemes because they lack a hostname and port. Instead, URIs like ``about:blank'' should inherit the origin of the document in which they are contained~\cite{mdn_origin}; since this URI inherits the same origin as its parent document, the same-origin policy dictates that a local frame will have access to the parent document.
There are other non-standard URIs in addition to ``about:blank'' for which content blockers may confuse the origin. The ``about'' prefix is intended to reference an application's internal resources, so other URIs starting with ``about'' besides ``about:blank'' (such as ``about:srcdoc'') could be misinterpreted~\cite{rfc6694}. There are also several other prefixes, such as ``blob'', ``file'', and ``data''~\cite{html}; in particular, the ``data'' prefix is supposed to receive an empty security context~\cite{mdn_origin}.  However, we find (see Section~\ref{sec:results:prevalence}) that these other URI prefixes make up only 0.5\% of local frames created by the websites we study.
%We focus on ``about:blank'' because of its prevalence and leave the investigation of other prefixes to future work.

\subsubsection{Common usage}

The lack of isolation between local frames and the main page has made them attractive for serving ads for over 15 years. 
In particular, local frames have been used to serve ``Rich Media'' ads, which have features such as displaying videos, expanding their size, and moving around the page~\cite{googlerichmedia,iab_rich_media}. These features require the ad to have access to the main page, so in 2008 the Interactive Advertising Bureau (IAB) published a set of best practices recommending websites to load ads in ``friendly iframes''---a local frame with source ``about:self''~\cite{iab_best_practices}.
% The guidelines state that serving ads in this manner allows the ads to access features (e.g. the JavaScript \texttt{document.write} API) which would otherwise interfere with the main page.
The Google Ad Manager documentation also notes that these friendly iframes are better able to collect metrics like viewability compared to cross-domain iframes~\cite{googleviewability}.
Given local frames' power to access the main page, online resources note that websites loading ads in these frames should have a trusted relationship with advertisers~\cite{mediumblog,humansecurity}.

Beyond ads, local frames are also used for tracking purposes. FingerprintJS~\cite{fpjs}, a popular open-source library for browser fingerprinting, creates a local frame when identifying the fonts installed on a given machine. The FingerprintJS documentation indicates that a local frame is used so that the fingerprinting script does not affect the main page and vice versa.
%\footnote{\url{https://github.com/fingerprintjs/fingerprintjs/blob/76fca7ac0e15a6ceb409faca497c6e8223d64991/src/sources/fonts.ts\#L72}}

% We investigated a popular open-source library for browser fingerprinting, FingerprintJS.\footnote{\url{https://github.com/fingerprintjs/fingerprintjs}} We found that FingerprintJS creates a local frame that includes 159 HTML text elements. FingerprintJS uses these text elements to render various fonts in order to determine what fonts the user has installed. In fact, the FingerprintJS documentation indicates that a local frame is used so that the fingerprinting does not affect the main page, and vice versa.\footnote{\url{https://github.com/fingerprintjs/fingerprintjs/blob/76fca7ac0e15a6ceb409faca497c6e8223d64991/src/sources/fonts.ts\#L72}}

% Another reason why local frames are used is that historically, local frames were the easiest way to create parallelism. \alisha{I'm having a hard time finding references for this}
% Historically, the blogosphere has recommend using iFrames to load ads---not just because iFrames can load third-party content, but also because iFrames load in parallel with the main page, and thus can improve the performance of rendering ads.

\subsection{Capabilities of Content Blockers}\label{sec:capabilities}

We consider four distinct capabilities of popular content blockers: request blocking, 
resource replacement, scriptlet injection, and cosmetic filtering.

\subsubsection{Request blocking}\label{sec:capabilities:req}

% \alisha{We should de-emphasize the third-party modifier}
The basic function of content blockers is to block network requests to undesirable targets such as ads and privacy-invasive tracking scripts. 
Network requests can either be created in the standard way (i.e., creating an HTML script or image element with the source set to the desired resource), or through asynchronous JavaScript APIs (e.g., \texttt{fetch} and \texttt{XMLHttpRequest} calls).
Sometimes outright blocking these requests can lead to site breakage. So, many content blockers have the option to block these requests only if they originate from a third-party context.
Another way to prevent breakage is to create fine-grained entries in the filter list that target specific scripts or paths; in this scenario, content blockers may allow a third-party iframe created by a third-party tracker to load, but block specific content within that iframe.
% For example, EasyList blocks third-party requests to common paths for displaying ads (for every website), such as \texttt{/promo.php} and \texttt{/a-ads}.
%
%To understand how widely this capability is used, we can check how many request blocking %rules there are in standard filter lists like EasyList and EasyPrivacy. 
As of August 17, 2024 EasyList contains 2,294 rules that use a third-party modifier to define the context in which these rules apply, and EasyPrivacy contains 4,151 rules with the modifier~\cite{easylist,easyprivacy}.

\subsubsection{Resource replacement}

Even if content blockers attempt to block only requests occurring in a third-party context, they can still trigger site breakage. In particular, some websites will expect to see the side effects of their scripts, such as defining certain functions and variables. If a content blocker prevents that script from executing, the website will not see the expected side effects and may throw errors. Alternatively, the website could detect a ``bait'' resource being blocked (without having to check for side effects) and thus determine that the user is using a content blocker~\cite{nithyanand2016adblocking}. So, many popular content blockers support redirecting scripts to benign versions that define the expected side effects while avoiding privacy-invasive behavior. For example, 
% tracking pixels can be redirected to empty images and scripts can be redirected to ``no-op'' scripts that run without executing any commands. 
uBlock Origin and AdGuard create benign versions of specific scripts that define expected objects (e.g., \texttt{window.google.ima} in the case of Google Ad Manager scripts\footnote{\url{https://github.com/gorhill/uBlock/blob/2c60b331e39e96114386e568d028240b37cdeefc/src/web_accessible_resources/google-ima.js\#L854}}).

\subsubsection{Scriptlet injection}

Beyond blocking and redirecting requests, some content blockers inject their own JavaScript into loaded webpages to provide more extensive protection from tracking. For example, content blockers cannot modify cookies through blocking network requests. Scriptlets, however, can be used to modify cookies and perform other tasks like disabling access to certain JavaScript APIs.
% However, content blockers like uBlock Origin, AdGuard, and Brave provide scriptlets that can modify cookies, as well as perform other functions like preventing scripts from reading or writing properties and disabling access to certain JavaScript APIs. 
For example, uBlock Origin uses scriptlets to remove telemetry cookies from Bing\footnote{\url{https://github.com/uBlockOrigin/uAssets/blob/e19390098b75344eaeaefcf33d599b840b75663d/filters/privacy.txt\#L649}},
prevent some websites from saving browser fingerprints\footnote{\url{https://github.com/uBlockOrigin/uAssets/blob/b38c34e9ff6f6467144954293100b166e36033e8/filters/filters-2024.txt\#L423}},
and disguise the use of the content blocker on streaming sites like \texttt{cbs.com} and \texttt{paramountplus.com}\footnote{\url{https://github.com/uBlockOrigin/uAssets/blob/b38c34e9ff6f6467144954293100b166e36033e8/filters/filters-2024.txt\#L92}}.

\subsubsection{Cosmetic filtering}

%\bingad

Finally, sometimes ads cannot be blocked by preventing network requests. Websites may display ads through inline HTML, meaning no network requests are used to render the ads. Alternatively, websites may use network requests to display ads, but include benign, functional code in the same script~\cite{amjad2021trackersift}; if content blockers were to block these scripts, they could break website functionality. Instead, blockers can hide ads through cosmetic filters (also known as ``element hiding'') that identify unwanted content through HTML and/or CSS selectors.
% \todo{@Pete do you know if/why this is true?}. 
% Many content blockers support selecting elements by their HTML and/or CSS selectors, and hiding these elements from users. 

%For example, uBlock Origin specifies six cosmetic filters to hide ads on \texttt{bing.com};\footnote{\url{https://github.com/uBlockOrigin/uAssets/blob/346c75cd7fd0fc7f8f8543a02b488a68fbf25c5a/filters/filters-2024.txt\#L201-L208}} we show the effect of one of these filters in Figure~\ref{fig:bingad}. As of August 17, 2024, EasyList has 22,097 cosmetic filtering rules~\cite{easylist}.

\subsection{Content Blocker Limitations}

There is a long-running arms race between website publishers who display ads to generate revenue, and content blockers that attempt to hide these ads and block tracking scripts to improve user privacy and user experience. Many publishers try to detect the use of content blockers and change their websites to discourage the use of these tools, a well-studied practice known as anti-adblocking~\cite{iqbal2017ad,mughees2017detecting,zhu2018measuring}. Studies disagree on the prevalence of anti-adblocking, reporting rates ranging from 0.7\%~\cite{mughees2017detecting} to 30.5\%~\cite{zhu2018measuring}.

Websites also try to evade content blockers by exploiting a widely acknowledged limitation of popular content blockers: the reliance on matching URLs to handcrafted filter lists.
In 2016 Wang \textit{et al.} proposed a system to allow Web publishers to evade content blockers by automatically randomizing URLs and HTML attributes~\cite{wang2016webranz}. While it is unclear if this particular system is used in practice, websites do change the way they host tracking content for evasion. 
A 2020 analysis of 10K websites found 1,612 instances of a blocked script being hosted on a new domain, as well as other techniques to change where tracking scripts are hosted~\cite{snyder2020filters}.
Our work similarly investigates the fragility of content blockers matching URLs against filter lists; however, we find evasion can take place without changing how tracking resources are hosted, by instead wrapping the request (or the requested content) inside a local frame.

Given the brittle nature of filter lists, alternative systems have been proposed to identify tracking resources. The AdGraph system generates a model of websites as graphs and feeds the graph context into an ML model to determine if resources are tracking or non-tracking~\cite{iqbal2020adgraph}.
% They find that the graph model helps identify tracking scripts through structural properties (e.g. node degree connectivity, attributes of parent nodes), resulting in high accuracy and even correcting mistakes in filter lists.
In 2021, Chen \textit{et al.} proposed using JavaScript event-loop signatures to identify tracking code~\cite{chen2021detecting}. They found 12.5\% of websites were able to evade filter lists by including scripts that contain tracking behavior but were not previously included in filter lists. 
However, a system like the one Chen \textit{et al.} propose would not solve the issues we identify with local frames, as we find tools fail to apply filter list rules to local frames, thus creating a vector for evasion.

Finally, content blockers implemented as browser extensions inherit the limitations of the browsers upon which they rely.
In 2016, Bashir \textit{et al.} measured the impact of a known bug in the Chrome WebSocket API that allowed websites to evade content blockers~\cite{bashir2018tracking}. 
% ; an inability to capture all requests can allow websites to evade these tools. In 2016 Bashir \textit{et al.} measured evasion of filter lists through a known bug in the Chrome WebSocket API that allowed requests to bypass content blockers~\cite{bashir2018tracking}. 
They found 2\% of websites used the WebSocket API, and 60\% of those sites were opening WebSockets to advertising and analytics companies.

%% file: content/crawl_analysis.tex
\section{Local Frame Usage}
\label{sec:crawl_results}

We motivate our study of vulnerable tools with
%In this Section we present 
a measurement of how frequently \LFs{} are used on the Web
currently. We also measure how often websites use \LFs{} to carry out the
kinds of behaviors that content-blocking tools target.
We note, however, that our measurement is not able to infer the intent
of website authors. Our results only show how often websites use \LFs{}
in ways that may circumvent existing privacy tools; we are not suggesting that website authors are using local frames with the sole intent of circumventing privacy tools.

We find that \LFs{} are widely used on the Web, appearing on more than half (\CrawlSitesWithFramesPct{}) of all websites we study. Furthermore,
websites frequently use \LFs{} to fingerprint users and make requests to 
URLs defined to be privacy harming (or otherwise unwanted) by
popular blocklists---behaviors that privacy and content-filtering Web tools target and modify.
% carry out the kinds of behaviors that
% privacy and content-filtering Web tools target and modify like making
% requests for URLs defined to be privacy harming (or otherwise unwanted) by
% popular blocklists and fingerprinting users.
%, executing \JS{}, and display HTML elements. Finally
% Interestingly, we find that 
% %not only do websites make many requests from within \LFs{}, but that 
% many of the websites against which privacy tools inject scriptlets
% (\IE{} custom-crafted scripts that modify a page's execution environment
% to prevent access to certain \JS{} APIs or properties) execute \JS{} in
% \LFs{}, circumventing---intentionally or otherwise---these protections.
% %intended by privacy tools.

\subsection{Measurement Methodology}
\label{sec:crawl_results:methodology}

We measure how, and how often, \LFs{} are used on popular websites in multiple
steps.

\subsubsection{Website selection}
\label{sec:crawl_results:methodology:sites}
We use the Tranco top-sites list to select websites with varying degrees of popularity at time of measurement. 
%We consider the top-15k sites to ensure that we measure \LF{} use on a large number of websites. We initially
%considered just the top-10k sites, but found that 
We collect data for
\CrawlNumSitesCompleted{} sites in total, including \CrawlNumSitesCompletedPopular{} of the Tranco top 15K, 5,000 websites uniformly sampled between the ranks of 15K to 100K, and 5,000 websites between ranks 100K and 1M.
% or \CrawlNumSitesCompletedPct{} of the Tranco top-15K list.
Our crawl attempts fail on \CrawlNumSitesUncompletedPct{}
of websites, either because the domains are not used to serve
websites (\EG{} CDN domains used to serve page assets), the domains are
geo-restricted, or, in a small number of cases, because of compatibility issues
with our crawling tools. 
Hence, successfully crawling 5K sites required trying  \CrawlNumSitesAttemptedMedium{} sites between 15K--100K and \CrawlNumSitesAttemptedUnpopular{} among the ranks 100K--1M.

\subsubsection{Tools and vantage points}
\label{sec:crawl_results:methodology:tools}

We visit each of the selected sites using
%a PageGraph\todo{CITE}
%based crawler. PageGraph is a Web measurement tool based on a current
%fork of Chromium 
the PageGraph crawler~\cite{pagegraph}, a Web-measurement tool based on a current
fork of Chromium. The crawler records 1) 
%an extremely broad  set of
salient
events that occur during the loading, rendering, and executing of a website
(\EG{} network requests issued; HTML elements created, modified, or inserted
in the page; WebAPIs executed by \JS{}, etc.), and 2) attribution
information for each of these events (\EG{} the HTML element or \JS{}
code unit responsible for a network request, whether an HTML element
was created as a result of parsing an HTML text or by a script, which script
called which WebAPI, etc.). 
%PageGraph (and the related tooling) 
The crawler outputs this
log of events and actors into a graph, yielding one graph per measured website.

We conduct our crawl 
%measured each of the \CrawlNumSitesAttempted{} selected websites (\IE{}
%the Tranco 15k) 
from AWS EC2 servers in Amazon's \texttt{us-west-2} region.
We configure the crawler to appear identical to a typical Chromium-based
browser by running the browser in ``head-full'' mode, removing \JS{}
properties that indicate to a website that a browser is a crawler (\EG{}
\texttt{window.webdriver}), and taking additional, similar steps to avoid signals that prior research
has identified can influence Web measurement results~\cite{jueckstock2021towards}.
For each website, the crawler visits the root page for each domain (\IE{}
either \texttt{https://domain.example} or \texttt{http://domain.example}),
waits \CrawlDwellTime{},
%to allow the page to fully render, 
and then 
%exports the PageGraph-generated graph.
%for each domain, yielding \CrawlNumSitesCompleted{} graphs.
records the resulting event graph.

\subsection{Results}
\label{sec:crawl_results:methodology:processing}

We post-process our crawl data
%We use existing tools designed to
%query information about PageGraph recordings\todo{CITE} 
to extract the
first-party and third-party \LFs{} from each website and analyze their behaviors.

\input{tables/frames_by_rank}

\input{tables/frames}
\input{tables/privacy-suspect-events}

\input{tables/third_party_frames}

\subsubsection{Local-frame prevalence}\label{sec:results:prevalence}

In each website graph, we identify frames that are local during the entire page execution. This
is an important filter because internally Chromium initializes all
child frames (\EG{} \texttt{<iframe>}, \texttt{<frame>}, \texttt{<object>}) as a \LF{}, even if that frame is then navigated to another URL.
%Our study only considers frames that remain local, and is scoped to considering frames with the URIs of ``about:blank'' and ``about:srcdoc''.
%   \item \textbf{Immediate first-party \LFs{}}: We only consider
%     frames that are an immediate child of the main, top-level frame
%     in the document (\IE{} the main HTML document). This means we do not
%     include \LFs{} that are children of third-party frames (\EG{} a \LF{}
%     that is included by an ad) or children of other \LFs{}, among other
%     examples.
% \end{enumerate}
We find that ``about:blank'' is by far the most common local-frame URI, accounting for 95.8\% of all local frames in our dataset; ``about:srcdoc'' (the only other URI with the ``about'' prefix we see) accounts for 3.7\% of local frames, followed by ``blob'' at 0.4\% and ``data'' at 0.1\%. The prevalence of each prefix is consistent across website ranks. Results reported in the remainder of this section pertain only to ``about:blank'' and ``about:srcdoc'' local frames.

% \noindent The first criteria ensures we consider only local frames.  The second focuses our study
% %We limit our measurement using the above criteria mainlyto focus 
% on choices directly attributable to the site
% being measured. While it might also be interesting to remove these filtering
% criteria and consider a broader range of \LFs{} (and their activites), such measurements are beyond the scope of this work.

%Local frames are extremely common on the Web:
%over half  of the \CrawlNumSitesCompleted{} measured
%sites include at least one \LF{}.
We find \CrawlNumFramesAllSites{} \LFs{} on
\CrawlSitesWithFramesNum{} distinct sites, and,
%(\CrawlSitesWithFramesPct{} percent of sites measured).
%Table \ref{table:frames} presents more details about
%the distribution of \LFs{} in our crawl results.
%
as shown in Table~\ref{tab:rank}, there is no obvious trend in local-frame prevalence across website popularity.  
%We find that websites in the rank interval {[}5K,10K) have a larger percentage of local frames than the other categories, but we are unable to explain why. In this table we also see that 
The vast majority of websites using local frames contain at least one first-party local frame; third-party local frames are significantly less common at all popularity ranks.

%As shown in Table~\ref{table:frames}, 
%Even given the filtering criteria used to limit the number of frames under
%consideration (as discussed in Section \ref{sec:crawl_results:methodology:processing}),
%Over half (\CrawlSitesWithFramesPct{}) of the \CrawlNumSitesCompleted{} measured
%sites include at least one \LF{}, with the sites
%including on average \CrawlSitesWithFramesMean{} \LFs{}.
% We also find that the ``about:blank'' URI is far more common than ``about:srcdoc'' (which accounts for only 4.3\% of the local frames we identify).
% Appendix~\ref{appendix:measurement} breaks down local-frame prevalence by website popularity.

\subsubsection{Privacy-relevant events}

From the \LFs{} we discover, we
extract instances of the following behaviors to understand how often \LFs{}
are used for the kinds of activities that are targeted by privacy tools:
\begin{enumerate}
\item Fingerprinting-related API calls including canvas, navigator, screen, and WebGL (the full list is in Appendix~\ref{appendix:fp_apis}),
  \item Requests made within a \LF{} (\EG{} images requested, \texttt{fetch} and
    \texttt{XMLHttpRequest} calls, etc.),
  \item Calls to privacy-relevant WebAPIs and \JS{} built-ins
  %, for the subset of these instrumented by PageGraph\todo{CITE}
  (\EG{} calls to
    \texttt{performance.now()}, Canvas APIs, etc.), and
  \item Non-default HTML elements included (\IE{} all HTML elements inserted
    into the frame except those created automatically such as \texttt{<html>}, \texttt{<head>}, \texttt{<body>},
    etc.).
\end{enumerate}

%We find that \LFs{} are frequently used to carry out the categories
%of behaviors that privacy tools block or modify. As presented in 
%The three right-most columns of 
Table~\ref{table:frames} shows that many sites use \LFs{}
to perform fingerprinting, make network requests, execute \JS{}, 
%(which we discuss in more detail in the next section), 
and present HTML elements to users.
%We note though that this table only identifies privacy-relevant behaviors, or
%behaviors that privacy tools block or modify. In other words, this table
Approximately a third (\CrawlFPAPIsSites{}) of all websites containing local frames use them to perform fingerprinting, a clear privacy concern.
%Unlike privacy-suspect requests, however, 
%(and we do not find any clear trends with regard to site popularity).
%trend: the percentage of sites fingerprinting in local frames is 19.6\%, 23.7\%, and 15.0\% in ranks [1-15K), [15K-100K), and [100K-1M), respectively.
Users may wish to manage the remaining behaviors
%presents the instances where sites use \LFs{} to carry out 
with privacy tools as well, and these tools may
be failing users by mishandling \LFs{}.

\subsubsection{Privacy-suspect events}

One particularly suspicious class of events are
%We filter
%the remaining events to a smaller subset of
%suspicious events, \IE{} instances where there is high chance
%that \LFs{} are carrying out specific activities that privacy tools
%would prevent or modify.
%, but may not be able to because of mishandling of
%\LFs{} in privacy tools.
%The first type of suspicious event we extract is
% \begin{enumerate}
%First, w
requests made by \LFs{} that would be blocked by popular filter
lists. We 
%identified these privacy-suspect requests by 
extract the requests
made from \LFs{} and use the adblock-rs library~\cite{adblockrs_code} to check
them against EasyList~\cite{easylist}, EasyPrivacy~\cite{easyprivacy}, and the additional
lists maintained by the uBlock Origin project~\cite{ublockorigin}. 
%However, we again note that we cannot determine if these behaviors are intentionally evading content blockers or not.
%
% %Second, w
% \item We identify \LFs{} that invoke privacy-relevant \JS{} calls (\EG{}
% WebAPI calls and \JS{} builtins), on pages where popular filter lists indicate
% \JS{} execution should be prevented or modified with a scriptlet. 
% %We note
% %again that our measurements are limited to the set of WebAPIs and \JS{} builtin
% %instrumented by the PageGraph\todo{remove reference} project---a large and significant %number of APIs,
% %but not all. 
% \end{enumerate}
%
%\input{tables/privacy-suspect-requests-receivers}
%
Table~\ref{table:privacy-suspect-events} presents the results of this analysis.
%Finally, w
We find that in a significant number of cases, sites use \LFs{}
to conduct specific behaviors that are very likely to be targeted for blocking
or modification by privacy tools. 
%Table \ref{table:privacy-suspect-events} lists these instances. 
We observe, for example, that \CrawlRequestsSitesLocalFrameBlockedPct{} of sites making requests inside of \LFs{} attempt to request content that should be blocked by popular filter lists.
% We also observe \CrawlJSCallsSuspectSitesNum{} sites where filter lists would modify the \JS{}
% execution environment, yet those sites execute \JS{} in \LFs{} (thus evading modification in tools that are vulnerable).
We find that the rate of privacy-suspect requests seems to increase slightly with popularity, which is consistent with prior work that reports popular websites perform more tracking~\cite{iqbal2021fingerprinting,englehardt2016online}.
%

%We report the top-10 entities (URLs are mapped to owning organization using the Disconnect entity list~\cite{disconnect}) that are targeted by these privacy-suspect requests in Table~\ref{tab:requests_receive}, all of which are advertising and analytics companies. We find that Google is the most common entity, contacted by 5$\times$ more sites than the next most-popular entity.

\input{tables/block_requests}
\subsubsection{Third-party local frames}

Third-party local frames are particularly intriguing, as they likely are not under direct control of the website publisher---who may not even be aware of their use.  We find \CrawlThirdPartyFramesTotal{} third-party local frames hosted on \CrawlSitesWithThirdPartyFramesTotal{} unique websites.
% On average, these sites contained 5.87 third-party \LFs{}, with the maximum being for a ``Twitter video downloader'' site with 53 third-party \LFs{}.
We map the security origin of the third-party \LF{} (i.e., the origin of the content loaded into the local frame) to its owning organization using the Disconnect entity list~\cite{disconnect}; if there is no entity for a given URL, we use the URL as the entity. 
For each popularity rank we report the top-10 entities that create third-party local frames by the number of sites in Table~\ref{tab:third_parties} (see Appendix~\ref{appendix:measurement} for the corresponding eTLD+1s).
Almost all entities are advertising and analytics companies.
Unsurprisingly, large analytic companies like Google, PubMatic, and Cloudflare appear in the top four entities for each popularity rank.
After that the entities become more varied, though the medium and unpopular ranks have high overlap.

%% file: tables/frames_by_rank.tex
\begin{table}[t]
\begin{tabular}{lcccc}
\toprule
\multicolumn{1}{c}{\multirow{2}{*}{Rank Interval}} & \multicolumn{1}{c}{\multirow{2}{*}{Sites Crawled}} & \multicolumn{3}{c}{Local Frame Prevalence}                                              \\
\multicolumn{2}{c}{}                               & \multicolumn{1}{c}{1p} & \multicolumn{1}{c}{3p} & \multicolumn{1}{c}{Either} \\
\midrule
{[1--5K)} & 3,815 & 52.9\% & 20.0\% & 56.3\% \\
{[5K--10K)} & 4,239 & 54.8\% & 25.2\% & 59.8\% \\
{[10K--15K)} & 3,911 & 47.6\% & 19.9\% & 51.0\% \\
{[15K--100K)} & 5,000 & 58.3\% & 23.6\% & 60.8\% \\
{[100K--1M)} & 5,000 & 46.8\% & 19.8\% & 50.3\% \\
\midrule
Overall & \CrawlNumSitesCompleted{} & 52.2\% & 21.7\% & 55.7\% \\
\bottomrule\\
\end{tabular}
\caption{Prevalence of local frames at various Tranco ranks.}
\vskip -2em
\label{tab:rank}
\end{table}

%% file: tables/frames.tex
\begin{table*}[t]
  \centering
    \begin{tabular}{lrrrrrr}
      \toprule
%                              & \multicolumn{2}{c}{Frames} \\
                              & 1p          & 3p    & \begin{tabular}[c]{@{}c@{}}Fingerprinting\\API Calls\end{tabular} & Requests               & JS/API Calls  & HTML \\
      \midrule
        \# Sites              & 7,942         &\CrawlSitesWithThirdPartyFramesTotal{}  & \CrawlFPAPIsSites{} & \CrawlRequestsSites{}  & \CrawlJSCallsSites{}  & 4,629 \\
        Mean                  & 2.07          & 1.31   & 1,900.97 & 4.38                   & 5,257.19                & 32.52 \\
        Median                & 1             & 0      & 0                      & 0                    & 0 & 0 \\
        Max                   & 98            & 76     & 327,600 & 431                    & 636,724               & 7,432 \\
        Total                 & 45,451        & 28,812 & 40,220,690 & \CrawlRequestsTotal{}  & \CrawlJSCallsTotal{}  & 733,813 \\
      \bottomrule\\
    \end{tabular}
  \caption{Statistics regarding the prevalence of various privacy-relevant behaviors occurring inside of \LFs{} created by the \CrawlNumSitesCompleted{}
    crawled websites.} 
    % The JS/API Calls and HTML count frames with \texttt{about:blank} URIs; other columns count all local frames.
  \label{table:frames}
  \vskip -2em
\end{table*}

%% file: tables/privacy-suspect-events.tex
\begin{table*}[t]
  \centering
    \begin{tabular}{lrr|rr|rr|rr} \hline
      % \toprule
      & \multicolumn{2}{c|}{Rank [1--15K)} & \multicolumn{2}{c|}{Rank [15K--100K)} & \multicolumn{2}{c|}{Rank [100K--1M)} & \multicolumn{2}{c}{Total} \\ \hline
    %    & \# & \% & \# & \% & \# & \% & \# & \% \\ \hline
      % \midrule
        \# Requests in dataset & 1,377,219 & 100.0\% & 579,966 & 100.0\% & 461,240 & 100.0\% & \CrawlRequestsAll{} & 100.0\% \\
        $\llcorner$ in a local frame & 56,280 & 4.1\% & 26,884 & 4.6\% & 13,098 & 2.8\% & \CrawlRequestsTotal{} & \CrawlRequestsTotalPct{} \\
        $\llcorner$ that should be blocked & 42,111 & 74.8\% & 19,679 & 73.2\% & 9,148 & 69.8\% & \CrawlRequestsLocalLocalFrameBlocked{} & \CrawlRequestsLocalLocalFrameBlockedPct{} \\ \hline
      % \midrule
        \# Sites crawled & \CrawlNumSitesCompletedPopular{} & 100.0\% & 5,000 & 100.0\% & 5,000 & 100.0\% & \CrawlNumSitesCompleted{} & 100.0\% \\
        $\llcorner$ making $>=1$ request & 10,863 & 90.8\% & 4,962 & 99.2\% & 4,971 & 99.4\% & \CrawlRequestsSitesAnyRequest{} & \CrawlRequestsSitesAnyRequestPct{} \\
        $\llcorner$ in a local frame & 2,833 & 26.1\% & 1,306 & 26.3\% & 939 & 18.9\% & \CrawlRequestsSites{} & \CrawlRequestsSitesPct{} \\
        $\llcorner$ that should be blocked & 1,887 & 66.6\% & \CrawlRequestsSitesLocalFrameBlockedMedium{} & \CrawlRequestsSitesLocalFrameBlockedMediumPct{} & \CrawlRequestsSitesLocalFrameBlockedUnpopular{} & \CrawlRequestsSitesLocalFrameBlockedUnpopularPct{} & \CrawlRequestsSitesLocalFrameBlocked{} & \CrawlRequestsSitesLocalFrameBlockedPct{} \\ \hline
      % \bottomrule
    \end{tabular}
    \vskip 1em
  \caption{Requests that sites make in local frames, including requests that
  popular Web security and privacy tools intend to block, but which
  are not blocked in some tools because of mishandling of frame origins.}
  \label{table:privacy-suspect-events}
  \vskip -1em
\end{table*}

%% file: tables/third_party_frames.tex
\begin{table*}[t]
    \centering
    \begin{tabular}{lrr|lrr|lrr} \hline
    % \toprule
    \multicolumn{3}{c|}{Rank [1--15K)} & \multicolumn{3}{c|}{Rank [15K--100K)} & \multicolumn{3}{c}{Rank [100K--1M)} \\
    Entity & \# Sites & \# Frames & Entity & \# Sites & \# Frames & Entity & \# Sites & \# Frames \\ \hline
    % \midrule
    Google & 1903 & 6369 & Google & 720 & 2531 & Google & 679 & 2541 \\
    PubMatic & 673 & 4314 & adtrafficquality.google & 556 & 691 & adtrafficquality.google & 374 & 398 \\
    Unity & 232 & 351 & PubMatic & 199 & 1049 & Cloudflare & 85 & 2124 \\
    Cloudflare & 213 & 1058 & Cloudflare & 108 & 2453& PubMatic & 78 & 413 \\
    Amazon & 172 & 303 & SeedTag & 42 & 219 & Amadeus & 19 & 20 \\
    Vidoomy & 52 & 113 & AdYouLike & 32 & 37 & SeedTag & 12 & 79 \\
    Datadome & 42 & 51 & admatic.de & 20 & 47 & Jivox & 8 & 16 \\
    NextMillennium & 36 & 89 & Amadeus & 16 & 19 & Yandex & 8 & 74 \\
    ConnectAdRealtime & 35 & 83 & Amazon & 15 & 21 & Chaturbate & 8 & 26 \\
    Piano & 33 & 136 & ConnectAdRealtime & 12 & 43 & AdYouLike & 6 & 6 \\ \hline
    % \bottomrule\\
    \end{tabular}
        \vskip 1em
    \caption{The top-10 entities of content loaded into third-party local frames sorted by the number of sites on which they appear.}
    %in and the number of local frames}
    \label{tab:third_parties}
    \vskip -1em
\end{table*}

%% file: tables/block_requests.tex
% Commands to revert later
% \renewcommand{\tablename}{Figure}
\small

\begin{table*}[t]
    \captionsetup{type=figure}
    \parbox{0.58\textwidth}{
        % All resources load
        \begin{tblr}{
                width=0.58\textwidth,
                colspec = {|X|X|X|X|X|X|X|X|X|X|X|X|},
                columns = {halign=c},
                column{1-2,4,7-8,10-12} = {wd=1pt},
                column{5-6} = {wd=0pt},
                vline{6} = {0pt},
                vline{1-5,7-13} = {0.5pt},
                colsep=1pt,
                rowsep=0.5pt,
                rows = {valign=m},
                row{1} = {1pt},
                row{2,5,8} = {belowsep=3pt,abovesep=2pt},
                row{3-4,6-7,9-10} = {1em},
                stretch = 0,
                hspan=minimal,
                % 1P script executes successfully
                cell{3}{1-6,7} = {bg=c1!35},
                cell{6}{2,8} = {bg=c1!35},
                cell{9}{3,9} = {bg=c1!35},
                % 3P script executes successfully
                cell{4}{1-6,7} = {bg=c2!35},
                cell{7}{2,8} = {bg=c2!35},
                cell{10}{3,9} = {bg=c2!35},
            }
            \SetHline{1-12}{0.5pt}
            \SetCell[c=12]{c} & 2 & 3 & 4 & 5 & 6 & 7 & 8 & 9 & 10 & 11 & 12 \\
            \SetHline{7-11}{0.5pt}
            \SetCell[c=5]{c} \textbf{First-Party Body} & 2 & 3 & 4 & 5 & & \SetCell[c=5]{c} \textbf{Third-Party iframe} & 2 & 3 & 4 & 5 & \\
            \SetCell[c=5]{c} {\texttt{firstparty.com} script executed} & 2 & 3 & 4 & 5 & & \SetCell[c=5]{c} {\texttt{firstparty.com} script executed} & 2 & 3 & 4 & 5 & \\
            \SetCell[c=5]{c} {\texttt{thirdparty.com} script executed} & 2 & 3 & 4 & 5 & & \SetCell[c=5]{c} {\texttt{thirdparty.com} script executed} & 2 & 3 & 4 & 5 & \\
            \SetHline{2-4,8-10}{0.5pt}
            & \SetCell[c=3]{c} \textbf{First-Party LF} & 3 & 4 & & & & \SetCell[c=3]{c} \textbf{Third-Party LF} & 3 & 4 & & \\
            & \SetCell[c=3]{c} {\texttt{firstparty.com} script executed} & 3 & 4 & & & & \SetCell[c=3]{c} {\texttt{firstparty.com} script executed} & 3 & 4 & & \\
            & \SetCell[c=3]{c} {\texttt{thirdparty.com} script executed} & 3 & 4 & & & & \SetCell[c=3]{c} {\texttt{thirdparty.com} script executed} & 3 & 4 & & \\
            \SetHline{3,9}{1pt}
            & & \textbf{First-Party Nested LF} & & & & & & \textbf{Third-Party Nested LF} & & & \\
            & & {\texttt{firstparty.com} script executed} & & & & & & {\texttt{firstparty.com} script executed} & & & \\
            & & {\texttt{thirdparty.com} script executed} & & & & & & {\texttt{thirdparty.com} script executed} & & & \\
            \SetHline{3,9}{1pt}
            & \SetCell[c=3]{c} & 2 & 3 & & & & \SetCell[c=3]{c} & 3 & 4 & & \\
            \SetHline{2-4,8-10}{0.5pt}
            \SetCell[c=5]{c} & 2 & 3 & 4 & 5 & & \SetCell[c=5]{c} & 2 & 3 & 4 & 5 & \\
            \SetHline{7-11}{0.5pt}
            \SetCell[c=12]{c} & 2 & 3 & 4 & 5 & 6 & 7 & 8 & 9 & 10 & 11 & 12 \\
            \SetHline{1-12}{0.5pt}
        \end{tblr}
        \subcaption{Structure of the test website when no filter list rules are applied, so all requests succeed.}
        \label{fig:blockreqs_unmodified}
    }
    \hfill
    \parbox{0.4\textwidth}{
        \begin{tblr}{
                width=0.4\textwidth,
                colspec = {|X|X|X|X|X|X|X|X|X|X|X|X|},
                columns = {halign=c},
                column{1-2,4,7-8,10-12} = {wd=1pt},
                column{5-6} = {wd=0pt},
                vline{6} = {0pt},
                vline{1-5,7-13} = {0.5pt},
                colsep=1pt,
                rowsep=0.5pt,
                rows = {valign=m},
                row{1} = {1pt},
                row{2,5,8} = {belowsep=3pt,abovesep=2pt},
                row{3-4,6-7,9-10} = {1em},
                stretch = 0,
                hspan=minimal,
            }
            \SetHline{1-12}{0.5pt}
            \SetCell[c=12]{c} & 2 & 3 & 4 & 5 & 6 & 7 & 8 & 9 & 10 & 11 & 12 \\
            \SetHline{7-11}{0.5pt}
            \SetCell[c=5]{c} \textbf{First-Party Body} & 2 & 3 & 4 & 5 & & \SetCell[c=5]{c} \textbf{Third-Party iframe} & 2 & 3 & 4 & 5 & \\
            \SetCell[c=5]{c} {[no text]} & 2 & 3 & 4 & 5 & & \SetCell[c=5]{c} {[no text]} & 2 & 3 & 4 & 5 & \\
            \SetCell[c=5]{c} {[no text]} & 2 & 3 & 4 & 5 & & \SetCell[c=5]{c} {[no text]} & 2 & 3 & 4 & 5 & \\
            \SetHline{2-4,8-10}{0.5pt}
            & \SetCell[c=3]{c} \textbf{First-Party LF} & 3 & 4 & & & & \SetCell[c=3]{c} \textbf{Third-Party LF} & 3 & 4 & & \\
            & \SetCell[c=3]{c} {[no text]} & 3 & 4 & & & & \SetCell[c=3]{c} {[no text]} & 3 & 4 & & \\
            & \SetCell[c=3]{c} {[no text]} & 3 & 4 & & & & \SetCell[c=3]{c} {[no text]} & 3 & 4 & & \\
            \SetHline{3,9}{1pt}
            & & \textbf{First-Party Nested LF} & & & & & & \textbf{Third-Party Nested LF} & & & \\
            & & {[no text]} & & & & & & {[no text]} & & & \\
            & & {[no text]} & & & & & & {[no text]} & & & \\
            \SetHline{3,9}{1pt}
            & \SetCell[c=3]{c} & 2 & 3 & & & & \SetCell[c=3]{c} & 3 & 4 & & \\
            \SetHline{2-4,8-10}{0.5pt}
            \SetCell[c=5]{c} & 2 & 3 & 4 & 5 & & \SetCell[c=5]{c} & 2 & 3 & 4 & 5 & \\
            \SetHline{7-11}{0.5pt}
            \SetCell[c=12]{c} & 2 & 3 & 4 & 5 & 6 & 7 & 8 & 9 & 10 & 11 & 12 \\
            \SetHline{1-12}{0.5pt}
        \end{tblr}
        \subcaption{Expected behavior with filter list rules to block all requests to the resources.}
        \label{fig:blockreqs_all}
    }
    \vskip -0.5em
    \caption{Structure of the test website for blocking requests and the expected behavior for RQ1. \textnormal{Cells highlighted in yellow indicate successful script requests from \texttt{firstparty.com} and cells highlighted in red indicate successful script requests from \texttt{thirdparty.com}.}}
\end{table*}
\vskip -1em

% Revert commands
\normalsize

%% file: content/vuln_tools.tex
\section{Vulnerable Content Blockers}\label{sec:tool_tests}

% As shown in Section~\ref{sec:crawl_results}, local frames are extremely prevalent on popular websites. 
In this section, we show that websites can exploit local frames to bypass the intended behavior of popular content blockers. We start by 
%defining the four core capabilities in popular content blockers. We then 
designing tests to determine whether each capability is executed correctly; we say a capability can be bypassed if a rule that is intended to be applied to a local frame is not applied. We also check if a capability is misapplied by checking if a rule that is not intended to be applied to a local frame is applied. We then discuss the tools we choose to study, and finally present our results. We find five major products (the Brave Browser, Safari Content Blocking, uBlock Origin Lite, AdBlock Plus, and AdGuard) have vulnerabilities that expose users to privacy-invasive tracking and visible ads, both of which are intended to be blocked.  One additional tool---DuckDuckGo---does not expose users but contains a vulnerability in its accounting functionality.
 
\subsection{Designing Tests for Capabilities}\label{sec:test_design}

For each capability we design test pages to determine if local frames can evade it. Each test uses two websites, one first-party and one third-party (referred to as \texttt{firstparty.com} and \texttt{thirdparty.com}), and we check whether the capability can be evaded in a first-party local frame and/or a third-party local frame.  For ease of implementation, our tests employ ``about:blank'' local frames; it is possible that more extensive testing with other local-frame URIs might uncover additional vulnerabilities.
% We plan to make our code for these test pages publicly available upon publication.
%
Our tests are not meant to be comprehensive for all functionality provided by content blockers, nor comprehensive of all possible code paths that implement each capability.
However, even our limited set of tests reveal mishandling of local frames in every tool we study.
Our test pages, and the filter list rules used for each tool, are publicly available at \texttt{https://osf.io/9yq57/files}.

%We also provide additional details on our tests (namely, testing local frames nested inside each other) in Appendix~\ref{appendix:testdetails}.

\begin{lstlisting}[language=HTML, deletekeywords={for, FRAME}, float, floatplacement=h, label=lst:teststructure, caption={Structure of a website \texttt{firstparty.com} with a first-party local frame, a nested first-party local frame, a third-party local frame, and a nested third-party local frame.\vspace{-2em}}]
 <body>
   <!-- Main frame for firstparty.com -->
   <iframe src="about:blank">
     <!-- Local frame for firstparty.com -->
     <iframe src="about:blank">
       <!-- Nested local frame for firstparty.com -->
     </iframe>
   </iframe>

   <iframe src="https://thirdparty.com">
     <!-- Main frame for thirdparty.com -->
     <iframe src="about:blank">
       <!-- Local frame for thirdparty.com -->
       <iframe src="about:blank">
        <!-- Nested local frame for thirdparty.com -->
       </iframe>
     </iframe>
   </iframe>
 </body>\end{lstlisting}
%
%\vskip 1em

\subsubsection{Request blocking}\label{sec:blockreqtestdesign}

For this test, we create a test website \texttt{firstparty.com} following the structure in Listing~\ref{lst:teststructure}, where all six (local and non-local) frames request two JavaScript files: one from \texttt{firstparty.com} and one from \texttt{thirdparty.com}. Both scripts create a text element inside the frames in which they are requested; the script from \texttt{firstparty.com} adds the text ``firstparty.com script executed'' while the script from \texttt{thirdparty.com} adds the text ``thirdparty.com script executed''. A representation of the resulting page (without any filter list rules applied) is shown in Figure~\ref{fig:blockreqs_unmodified}. 

% This website contains two nested first-party local frames, as well as a third-party iFrame with source \texttt{thirdparty.com}. Inside that third-party iFrame, there are two nested local frames. Each frame requests a script from \texttt{firstparty.com}, which adds the text "First Party Local Frame." Each frame also requests a script from \texttt{thirdparty.com}, which adds the text "Third Party Local Frame."

Concretely, we test \textbf{RQ1}: If a resource is supposed to be blocked, can it be loaded inside a local frame? If a content-blocking tool blocks both scripts from being requested by any domain, then neither script should get executed in any frame, and thus no text should be added to the website; this expected behavior is shown in Figure~\ref{fig:blockreqs_all}.  We consider more nuanced variants (i.e., if the script executes in only first- or third-party contexts) in Appendix~\ref{appendix:blockreq}, but our tests reveal that tools 
% (DuckDuckGo being the exception) -- the DDG nuance is that it's vulnerable in nested local frames. They don't have an issue with partyness
either handle all cases correctly or none.% (in the case of Brave iOS).

\input{tables/ajax}
\input{tables/scriptlet_and_cosmetic}

\subsubsection{Resource replacement}

There are many resources that popular content blockers can redirect; we focus our testing on AJAX requests as a representative example.  (We suspect the content-blocker code paths that compute request origins are likely the same no matter which type of resource is redirected.) AJAX requests are asynchronous \texttt{XMLHttpRequests} made to another webserver. On our webpage, we include two \texttt{h1} (header) elements in every frame. Each frame also includes JavaScript code that replaces the first \texttt{h1} element with the contents of a text file retrieved from \texttt{firstparty.com} (which reads ``firstparty.com AJAX executed''), and replaces the second \texttt{h1} element with the contents of a text file from \texttt{thirdparty.com} (``thirdparty.com AJAX executed'').

We show a representation of our test website without any AJAX requests made in Figure~\ref{fig:ajax_before}, and with all AJAX requests completed in Figure~\ref{fig:ajax_after}. In our test, we consider \textbf{RQ2}: If we define a filter list rule to redirect one of the text files fetched in the AJAX request to an empty file, does the original AJAX request still occur? In Figure~\ref{fig:ajax_3p}, we show an example of the expected behavior for redirecting requests to the text file hosted on \texttt{thirdparty.com}. We also test redirecting requests for \texttt{firstparty.com} (not shown).

\subsubsection{Scriptlet injection}

To test if scriptlet injection can be evaded by local frames, we create two filter-list rules, each employing a distinct ``set-constant'' scriptlet. The ``set-constant'' scriptlets define a new property, \texttt{scriptletvalue}, of the \texttt{window} object for a given domain. Both of the scriptlets set values for the \texttt{window.scriptletvalue} object, but to different constants: 1 for first-party and 42 for third-party websites. By using different values, we can check if a third-party local frame is confused for a first-party frame or vice versa. Concretely, we evaluate \textbf{RQ3}: For each local frame, is the correct scriptlet injected? 
%value of \texttt{window.scriptletvalue} read?
In our test website, each frame creates a text element on the page reporting the value of \texttt{window.scriptletvalue}. 
%We show the expected behavior in a representation of the website in Figure~\ref{fig:scriptlet_expected}; the first-party body and frames all report the value of \texttt{window.scriptletvalue} as 1, and the third-party frames all report the value as 42.

\subsubsection{Cosmetic filtering}

Our test for cosmetic filtering checks if elements can be selected and hidden properly. In our test website we create an \texttt{h1} element with the class \texttt{cosmetic-filter} in every frame, as shown in Figure~\ref{fig:cosmetic_unmodified}. We then test \textbf{RQ4}: If we define a rule to hide \texttt{h1} elements with the \texttt{cosmetic-filter} class, are they hidden? Figure~\ref{fig:cosmetic_expected} shows the expected behavior for hiding these elements on \texttt{thirdparty.com}; we also test the first-party website. %(not shown).

\subsection{Tools Studied}\label{sec:tools_studied}

Users seeking to protect their privacy typically install browser extensions that can intercept requests to and from trackers, or use browsers with strong privacy protections. 
% Some companies (notably DuckDuckGo) offer both browser extensions and a custom browser. 
We test popular browsers and browser extensions from six different organizations.
We focus on tools that employ filter lists because their behavior is easy to manipulate---i.e., by modifying the contents of the filter list---but
%they’re well-studied and simple to control the behavior of, 
we expect the unintuitive nature of local frames is equally likely to confound developers of any tool that needs to determine the party responsible for the content and behavior of a frame, regardless of the tool's blocking strategy.
%could be confused by the unintuitive behavior of local frames.
%We describe each below and provide additional details on the popularity and usage of the tools in Appendix~\ref{appendix:tools}.

\subsubsection{Extensions}

For browser extensions, we consider the most-popular Chrome and Firefox extensions.  Firefox recommends certain add-ons in the ``Privacy \& Security'' category~\cite{firefox_addons}. We select the top three (by download count) that rely on URL-based filter list rules: uBlock Origin, AdGuard, and DuckDuckGo. We do not test the two other tools in the top five; we exclude Ghostery because it blocks at the granularity of URL parameters and PrivacyBadger because it uses heuristics to determine what to block instead of a fixed set of filter list rules.
The same three extensions appear in the top five in the ``Privacy and Security'' category of the Chrome Web Store~\cite{chrome_web_store} using the default sorting method of ``Most relevant''. The other two tools in the top five are OnlineSecurity, a malware-defense system and Authenticator, a multifactor-authentication app. 
We also test AdBlockPlus, which, while not recommended by Firefox, has over 3 million users---second only to uBlock Origin. 
(The Chrome Web Store categorizes AdBlock Plus as a ``Workflow \& Planning'' extension; it does not appear in ``Privacy and Security''.)
We test these four extensions on both Firefox and Chome, as well as their iOS, Android, and Desktop apps where available.

We also checked the list of most-popular extensions in the Firefox Public Data Report, which reports metrics from Firefox desktop users each week~\cite{firefox_dashboard}. As of December 30, 2024, this report had three of the content blockers we study in the top-10 most-popular Add-ons, as well as one content blocker we do not study. The report lists uBlock Origin as the most popular extension, with a usage of 8.49\%. The next-most-popular content blocker (and the 4th-most-popular Add-on) was AdBlocker Ultimate, which primarily uses filter lists based on EasyList/Easyprivacy and AdGuard. The next-most-popular content blockers (ranked 7th and 8th respectively) are AdBlockPlus and DuckDuckGo.  (PrivacyBadger is ranked 9th overall, but as noted, uses heuristics instead of a filter list.)

We run browser extensions on Firefox version 129.0.1 or Chromium version 126.0.6478.182. We test Android apps on an emulated Pixel 8 Pro phone running API 35, and iOS apps on an iPhone 15 Pro running iOS version 17.5.1.

\textbf{uBlock Origin} is a widely used browser extension, with over 35 million downloads from the Chrome Web Store~\cite{ubo-cws}. uBlock Origin supports all four capabilities and offers both a Firefox and Chrome extension (we test version 1.59.0 of both).

Despite uBlock Origin's popularity, its Chrome extension is built with the Chrome Manifest V2 (MV2) framework, which Chrome has deprecated as of June 2025~\cite{chromemv2}; the details of MV2 or its successor, Manifest V3 (MV3), are not important for understanding this work, so we omit them. Though uBlock Origin is no longer supported due to MV2 deprecation, the maintainers now offer the uBlock Origin Lite extension, which is built with the new MV3 framework; we test version 2024.8.12.902. As of January 2025, uBlock Origin Lite had 1 million users on the Chrome Web Store~\cite{ubl-cws}.

\textbf{AdGuard} is another popular browser extension with over 14 million downloads from the Chrome Web Store as of January 2025~\cite{adg-cws}. We test the AdGuard Chrome extension (version 4.3.53) and Firefox extension (version 4.3.64), both of which offer all four capabilities. AdGuard also offers an iOS app that changes the user's browsing experience in the Safari mobile browser. At the time of writing, AdGuard claims to offer all capabilities except resource replacement for their iOS app~\cite{adguardfilters}. However, we are unable to get scriptlet injection rules working in the iOS app, so we only test request blocking and cosmetic filters.

\begin{lstlisting}[language=HTML, deletekeywords={for, FRAME}, float, floatplacement=h, label=lst:ddgstructure, caption={Structure of a website \texttt{firstparty.com} that embeds \texttt{intermediate.com} instead of directly embedding \texttt{thirdparty.com}. For clarity, we do not show nested local frames or requests made outside of local frames.\vspace{-2em}}]
<body>
  <iframe src="about:blank">
    <!-- Local frame for firstparty.com. Origin should be firstparty.com -->
    <script src="http://firstparty.com/should_be_allowed.js"></script>
    <script src="http://thirdparty.com/should_be_blocked.js"></script>
  </iframe>

  <iframe src="https://intermediate.com">
    <iframe src="about:blank">
      <!-- Local frame for intermediate.com. Origin should be intermediate.com -->
      <script src="http://firstparty.com/should_be_allowed.js"></script>
      <script src="http://thirdparty.com/should_be_blocked.js"></script>
    </iframe>
  </iframe>
</body>
\end{lstlisting}

\textbf{DuckDuckGo} is primarily known as a privacy-protecting search engine, but also offers browser extensions that block requests to trackers, as well as a standalone browser. All of the extensions and browser versions offer two of the capabilities outlined in Section~\ref{sec:capabilities}: request blocking and resource replacement. In particular, the latter is implemented through ``surrogate'' rules, which redirect specific JavaScript files into benign or no-op versions. We test the Chrome and Firefox extensions (versions 2024.7.10 and 2024.7.24 respectively) as well as the MacOS browser (version 1.101.0), iOS browser (version 7.134.0.0), and Android browser (version 5.210.2).

% For all of these apps, DuckDuckGo does not allow users to add custom filtering rules. Instead, we spoof DNS responses (through manipulating the \texttt{/etc/hosts} file) to replace \texttt{thirdparty.com} with domains that are blocked or redirected by DuckDuckGo's existing blocklists.\footnote{\url{https://github.com/duckduckgo/tracker-blocklists}} In particular, for request blocking we replace \texttt{thirdparty.com} with \texttt{doubleclick.net}. (The third-party scripts are still under our control, but just appear to be coming from \texttt{doubleclick.net}.) 

% One challenge we encounter is that DuckDuckGo's request blocking does not offer the third-party specifier option; requests are universally allowed, blocked, or redirected. So, we can only test RQ1.1 (blocking requests) but not RQ1.2 or RQ1.3 (blocking third-party and first-party requests, respectively).
One challenge we encounter is that we cannot properly test RQ1 using the setup in Listing~\ref{lst:teststructure}. DuckDuckGo does not allow users to define custom filter list rules, so (as we explain later in Section~\ref{sec:invoke}) we spoof the DNS response for \texttt{thirdparty.com} to match a common tracker. However, DuckDuckGo then blocks the entire third-party iframe for matching a known tracker, and so we cannot test the behavior of requests made inside the third-party iframe. 
% , since we want to test requests made inside third-party iframes, and DuckDuckGo blocks the iframe with source \texttt{thirdparty.com}. In other words, one cannot test requests for blocked content from inside an iframe with the same source because the iFrame itself will not be loaded if the domain is blocked.
As a workaround, we create another website, \texttt{intermediate.com}, that makes requests to \texttt{thirdparty.com} and change the source of the third-party iframe on \texttt{firstparty.com} from \texttt{thirdparty.com} to \texttt{intermediate.com}; a simplified version of the resulting page is shown in Listing~\ref{lst:ddgstructure}. This approach enables us to make third-party requests to a resource that should be blocked or redirected.

\textbf{AdBlock Plus} is one of the most-popular privacy-focused browser extensions, with over 41 million downloads from the Chrome Web Store as of January 2025~\cite{abp-cws}. AdBlock Plus supports all four of the capabilities outlined in Section~\ref{sec:capabilities}. We test the AdBlock Plus Chrome extension (version 4.10.1), Firefox extension (version 4.5), and iOS app (version 2.2.16).

% We believe that the AdBlock Plus iOS app only supports request blocking and cosmetic filters. Based on the documentation for AdBlock Plus~\cite{abp_filterlists} we believe that the iOS app only loads in filter list rules from EasyList (and optionally the Acceptable Ads allowlist, which we disable for testing). EasyList only contains request blocking and cosmetic filter rules, and therefore we only test these capabilities for the iOS app. In order to conduct these tests, we modified the request blocking and cosmetic filter tests to match existing filters in EasyList; in particular, we check if requests to \texttt{doubleclick.net} are blocked, and change the cosmetic filter class from \texttt{cosmetic-filter} to \texttt{ADBAR}.

\subsubsection{Browsers and APIs}

Among privacy-focused browsers, we test the Brave Browser as it has 75.9 million monthly active users as of November 2024~\cite{brave-users}, and the Safari Content Blocker API because it powers the iOS filtering system for at least two of the tools we study (AdGuard and Brave), and its API may be relied upon by the billions of people who use Safari~\cite{safariusers}. 

\textbf{Safari Content Blocker.} Apple allows developers to create content-blocking extensions that modify Safari on macOS and iOS~\cite{safaricontentblocker}. Safari Content Blockers support only two of the capabilities in Section~\ref{sec:capabilities}: request blocking (with third-party modifiers) and cosmetic filtering. To test this native content-blocking functionality, we implement our own content-blocking extension 
%(which we will release upon publication) 
and test it on Safari (version 17.5) on MacOS 14.5; the code for these extensions is publicly available at \texttt{https://osf.io/9yq57/files}. 

% easyprivacy rules: 52704
% easylist rules: 71227
A Safari Content Blocker extension can only support 150,000 rules~\cite{safari_filter_limit}; while this number may seem large, the combined EasyList and EasyPrivacy lists (as of August 17, 2024) have 123,931 rules~\cite{easylist,easyprivacy}, and so any iOS apps wishing to include their own rules in addition to these standard lists are running out of space. This limitation impacts any iOS apps that implement request blocking and cosmetic filtering through the Safari content-blocking system, including the AdGuard and Brave iOS apps (see Section~\ref{sec:diss}).

\textbf{Brave Browser.} Brave is a browser with many privacy-enhancing features and supports all of the four capabilities previously outlined in Section~\ref{sec:capabilities}. We test the Brave MacOS app (version 1.68.141), iOS app (version 1.68), and Android app (version 1.68.137).

Through source-code analysis, we discover that there are two separate code paths for request blocking. Based on this discovery, we create two versions of our request-blocking test setup. The first version is exactly as described in Section~\ref{sec:blockreqtestdesign}. The second version makes AJAX requests instead of directly including the resources. 
While similar to our resource-replacement test, it seeks to test whether filter-list rules can block these requests, not (just) redirect them.
In this modified version of the test we rename the requested resource to match one of Brave's existing filter-list rules: 
%Instead of requesting \texttt{thirdparty.com/script.js}, 
we request \texttt{thirdparty.com/ads/index.js} to match the existing rule for \texttt{/ads/index}.

% \textbf{Firefox Enhanced Tracking Protection.} The last tool we test is Firefox's Enhanced Tracking Protection (ETP) feature. While this feature offers many privacy protections, it implements only one of the capabilities we consider: request blocking. 
% % In particular, ETP will block requests made to the Disconnect blocklist.\footnote{\url{https://github.com/mozilla-services/shavar-prod-lists/blob/128.0/disconnect-blacklist.json}} Like DuckDuckGo, Firefox does not support a third-party modifier, nor does it allow custom filters. We again modify our \texttt{/etc/hosts} file to map \texttt{thirdparty.com} to \texttt{doubleclick.net} for the request blocking test. 
% We test ETP in ``strict'' mode, which applies request blocking to all websites; the default mode only applies request blocking to incognito windows.

\input{tables/tool_results}

\subsection{Invoking Tests}\label{sec:invoke}

Our tests, as described in Section~\ref{sec:test_design}, require custom filter-list rules to be added to content blockers. However, not all content blockers allow users to add these custom rules. Concretely, we cannot add custom filter list rules to the AdBlock Plus iOS app (which loads EasyList rules~\cite{abp_filterlists}) and all DuckDuckGo browsers and browser extensions (which load a custom, publicly available set of blocklists~\cite{ddg_blocklist_ios,ddg_blocklist_android,ddg_blocklist_mac,ddg_blocklist_extension}).
% , and Firefox Enhanced Tracking Protection (which loads the Disconnect blocklist\footnote{\url{https://github.com/mozilla-services/shavar-prod-lists/blob/128.0/disconnect-blacklist.json}}).
%
For these tools, we instead modify our tests to match the tools' existing filter list rules. 
In order to match existing rules, we spoof DNS responses to map the hosts in those rules to our test websites. For browsers and browser extensions (i.e. DuckDuckGo), we modify our local \texttt{/etc/hosts} file. For iOS apps (i.e. AdBlock Plus and DuckDuckGo), we set up an HTTP proxy with the Charles Web-proxy tool~\cite{charles}. 

Specifically, for request-blocking tests, we map our third-party server to \texttt{doubleclick.net}, since this domain is blocked by all of the aforementioned filter lists.
% \texttt{thirdparty.com} to be \texttt{npttech.com/advertising.js}; according to DuckDuckGo's publicly available blocklists, this particular script should be redirected to \texttt{noop.js} on all platforms we test~\cite{ddg_blocklist_ios,ddg_blocklist_android,ddg_blocklist_mac,ddg_blocklist_extension}
To test resource redirection for DuckDuckGo, we replace our AJAX requests with standard Web requests to a particular script (\texttt{npttech.com/advertising.js}) that DuckDuckGo redirects to an empty script on all domains.
Finally, to test cosmetic filters in the AdBlock Plus iOS tool, we change the HTML class of the \texttt{h1} elements from \texttt{cosmetic-filter} to \texttt{ADBAR}.

In total, we create seven unique tests; one for each research question, as well as two variants of RQ1 (described in Appendix~\ref{appendix:blockreq}) and the tool-specific tests for Brave's alternative request-blocking code path and DuckDuckGo.

\subsection{Results}

In this section we discuss the results of our testing; Table~\ref{tab:vuln-results} presents a summary of the vulnerabilities we discover.  In some instances our initial findings led us to analyze the tools' source code.
%, uncovering additional concerns.

\subsubsection{AdBlock Plus}

We find that the Chrome and Firefox extensions of AdBlock Plus are not deterministically vulnerable to any of our tests. However, we discover that there is a race condition in the Chrome extension where local frames can load before the extension has an opportunity to inject scriptlets into them. (Fortunately, when a scriptlet wins the race and is injected, we find that the correct scriptlet is used and the local frames display the expected behavior.)

We also find that the AdBlock Plus iOS app does not inject cosmetic filters into local frames, which allows websites to show ads and unwanted content despite AdBlock Plus intending to block this content. It is possible that this is because AdBlock Plus implements cosmetic filtering through the Safari content blocking tool (see Section~\ref{sec:results:safari}),
%(which similarly does not inject cosmetic rules into local frames), 
but we are unable to verify this as the source code to AdBlock Plus is not publicly accessible.

\subsubsection{uBlock Origin}

% Aside from the same race condition observed in AdBlock Plus, the Chrome and Firefox extensions of uBlock Origin are not vulnerable to any of our tests.
Like AdBlock Plus, the uBlock Origin Chrome extension inconsistently applies scriptlets. When scriptlets are injected, we find that local frames display the expected behavior.

uBlock Origin Lite, on the other hand, does not inject any cosmetic filters into local frames. This allows websites to display ads that should otherwise be hidden, simply by putting them into a local frame.  The maintainers do not plan to fix this issue for now, as they believe that patching the issue could incur high performance overhead~\cite{ubo_disclousre}.  However, during our disclosure process, the maintainers of uBlock Origin Lite discovered that scriptlets were also not being injected into local frames and patched the issue.

\subsubsection{AdGuard}

Across all tested platforms, we find that AdGuard incorrectly computes the origin of third-party local frames for scriptlet injection and cosmetic filters. Specifically, AdGuard determines the origin of these frames to be the origin of the first-party website. In our tests, we find that scriptlet and cosmetic rules intended for the first-party website are applied to third-party local frames. Conversely, the scriptlet and cosmetic rules intended for the third-party website are not applied to third-party local frames.
Because of this issue, third-party content can easily evade scriptlet injection and cosmetic filtering. As described in Section~\ref{sec:capabilities}, scriptlets and cosmetic filters are used to block tracking scripts, disguise the use of content blocking tools, and hide ads. AdGuard users are subject to all of these consequences for third-party content.

Another consequence of mis-attributing the origin of local frames is that AdGuard can introduce site breakage by applying the rules for the first-party website to the third-party local frames. Scriptlets that modify the JavaScript API may have unintended consequences. For example, AdGuard's scriptlets can prevent websites from making network requests via the \texttt{fetch} API, or prevent event listeners from being added to elements, both of which are standard tools for Web development~\cite{adguardscriptlets}. Disabling these features could break the behavior of third-party content.

\subsubsection{Brave}

We find some of Brave's tools to be vulnerable to evasion of all four capabilities outlined in Section~\ref{sec:capabilities}. 
Across all tested platforms, scriptlets and cosmetic filters are not injected into local frames. The Brave iOS app can be evaded in another way, as resource replacement rules are not injected inside local frames. We find that AJAX requests made inside local frames are not redirected to empty text responses. While we only test AJAX requests, resource replacement is often used to redirect privacy-invasive scripts to benign versions. Users of Brave's iOS app can be subject to these invasive tracking scripts if websites simply make these requests from inside local frames.

More critically, we find that Brave's iOS app is vulnerable to evasion in request blocking. As described above, Brave implements two different code paths for request blocking, and we design a version of the request blocking test that makes AJAX requests and matches Brave's default filter list rules (i.e. not added by the user). This version of our test shows that local frames are able to make requests that should otherwise be blocked. The reason that Brave has two different code paths is because of the aforementioned limitations on the number of rules that can be loaded in a Safari Content Blocking extension. Brave compiles a subset of EasyList and EasyPrivacy, called the ``slim list'', that are loaded into the Safari Content Blocking extension~\cite{brave_slim}. However, Brave still attempts to block a larger set of requests than this subset. They cannot block any more standard Web requests because iOS restricts access to these requests~\cite{safaricontentblocker}. But apps still have access to asynchronous requests made through JavaScript APIs, such as AJAX requests and the \texttt{fetch} API. Brave attempts to block these asynchronous requests, but incorrectly computes the origin of the request originator\footnote{\url{https://github.com/brave/brave-core/blob/094608dbd95704ae314acbb9e05080c566afa5ad/ios/brave-ios/Sources/Brave/Frontend/UserContent/UserScripts/Scripts_Dynamic/Scripts/Paged/RequestBlockingScript.js\#L25}}. This mistake allows local frames to make requests to trackers (that are not in the aforementioned slim list).

% \todo{@Pete can you add more details: what set of URLs are vulnerable? Why doesn't this work for custom rules?}

% More critically, source code analysis shows that Brave's iOS app will load certain requests in local frames that should be blocked if they match particular patterns.  Hence we rename the third-party script in our test to match one such instance (renaming \texttt{thirdparty.com/script.js} to \texttt{thirdparty.com/ads/index.js} to match the existing rule \texttt{/ads/index}). While the request made by the main first-party body is correctly blocked, requests made by the first-party local frames are allowed. \todo{@Pete can you add more details: what set of URLs are vulnerable? Why doesn't this work for custom rules?} \todo{@Alisha: emphasize that the request blocking issue is in part because of limitations in Safari content blocking, forcing developers to add complexity to their codebase to get around these limitations, making it an ecosystem issue}

\subsubsection{DuckDuckGo}\label{sec:results:ddg}

We find that DuckDuckGo correctly blocks requests and redirects resources due to its block-by-default behavior; requests to trackers are blocked unless an exception is defined in an allowlist.
However, source-code analysis reveals a minor bug that demonstrates how common it is for security/privacy tools to mishandle local frames: The bug causes DuckDuckGo to mis-report how many trackers have been blocked from a given website. This error may mislead DuckDuckGo users attempting to understand the privacy-harming behaviors of the websites they visit.  Specifically, we find that requests blocked inside nested local frames are not properly accounted (see Appendix~\ref{appendix:testdetails}).

\input{tables/safari}
\input{tables/disclosure}

\subsubsection{Safari Content Blocker}\label{sec:results:safari}

We find that Safari Content Blockers do not inject cosmetic filters into any local frames, meaning Safari users may see ads that should be blocked. We emphasize that this is not just an issue for Safari users, but also for users of iOS apps that rely on Safari's content blocking functionality.

In addition, we find a larger discrepancy (though not a vulnerability) between how request blocking works on Safari and how request blocking is implemented by all the other content-blocking tools we study. In particular, the discrepancy is in computing if a request is made in a third-party context. Most tools have the behavior shown in Figure~\ref{fig:blockreqs_3p}; in particular, these tools allow requests to \texttt{thirdparty.com} from the third-party iframe. These tools consider these requests to be executing in a first-party context, because the origin of the request is the same as the origin of the document in which the request is made. Safari, on the other hand, has the behavior shown in Figure~\ref{fig:safari_blockreq}. Safari blocks the requests to \texttt{thirdparty.com} from the third-party iframe, meaning they compare all requests to the origin of the base webpage.
%the user is visiting.

This mismatch in computing ``partyness'' does not result in a vulnerability in this instance, but can lead to unexpected behavior. In particular, if the iOS apps that use Safari's content blocking functionality use the same filter lists used on other platforms, they will see different behavior. We disclosed this finding to Apple.
% in addition to the 19 local frame vulnerabilities.

%% file: tables/ajax.tex
\renewcommand{\tablename}{Figure}
\small

\begin{table*}[t]
    \captionsetup{type=figure}
    \parbox{0.32\textwidth}{
        % Timestamp 0: before requests execute
        \begin{tblr}{
                width=0.32\textwidth,
                colspec = {|X|X|X|X|X|X|X|X|X|X|X|X|},
                columns = {halign=c},
                column{1-2,4,7-8,10-12} = {wd=1pt},
                column{5-6} = {wd=0pt},
                vline{6} = {0pt},
                vline{1-5,7-13} = {0.5pt},
                colsep=1pt,
                rowsep=0.5pt,
                row{1} = {1pt},
                row{3-4,6-7,9-10} = {2.5em},
                row{2,5,8} = {belowsep=3pt,abovesep=2pt},
                stretch = 0,
                hspan=minimal,
            }
            \SetHline{1-12}{0.5pt}
            \SetCell[c=12]{c} & 2 & 3 & 4 & 5 & 6 & 7 & 8 & 9 & 10 & 11 & 12 \\
            \SetHline{7-11}{0.5pt}
            \SetCell[c=5]{c} \textbf{First-Party Body} & 2 & 3 & 4 & 5 & & \SetCell[c=5]{c} \textbf{Third-Party iframe} & 2 & 3 & 4 & 5 & \\
            \SetCell[c=5]{c} {Text to be \\[-1pt] replaced} & 2 & 3 & 4 & 5 & & \SetCell[c=5]{c} {Text to be \\[-1pt] replaced} & 2 & 3 & 4 & 5 & \\
            \SetCell[c=5]{c} {Text to be \\[-1pt] replaced} & 2 & 3 & 4 & 5 & & \SetCell[c=5]{c} {Text to be \\[-1pt] replaced} & 2 & 3 & 4 & 5 & \\
            \SetHline{2-4,8-10}{0.5pt}
            & \SetCell[c=3]{c} \textbf{First-Party LF} & 3 & 4 & & & & \SetCell[c=3]{c} \textbf{Third-Party LF} & 3 & 4 & & \\
            & \SetCell[c=3]{c} {Text to be \\[-1pt] replaced} & 3 & 4 & & & & \SetCell[c=3]{c} {Text to be \\[-1pt] replaced} & 3 & 4 & & \\
            & \SetCell[c=3]{c} {Text to be \\[-1pt] replaced} & 3 & 4 & & & & \SetCell[c=3]{c} {Text to be \\[-1pt] replaced} & 3 & 4 & & \\
            \SetHline{3,9}{1pt}
            & & \textbf{First-Party \\[-1pt] Nested LF} & & & & & & \textbf{Third-Party \\[-1pt] Nested LF} & & & \\
            & & {Text to be \\[-1pt] replaced} & & & & & & {Text to be \\[-1pt] replaced} & & & \\
            & & {Text to be \\[-1pt] replaced} & & & & & & {Text to be \\[-1pt] replaced} & & & \\
            \SetHline{3,9}{1pt}
            & \SetCell[c=3]{c} & 2 & 3 & & & & \SetCell[c=3]{c} & 3 & 4 & & \\
            \SetHline{2-4,8-10}{0.5pt}
            \SetCell[c=5]{c} & 2 & 3 & 4 & 5 & & \SetCell[c=5]{c} & 2 & 3 & 4 & 5 & \\
            \SetHline{7-11}{0.5pt}
            \SetCell[c=12]{c} & 2 & 3 & 4 & 5 & 6 & 7 & 8 & 9 & 10 & 11 & 12 \\
            \SetHline{1-12}{0.5pt}
        \end{tblr}
        \subcaption{Timestamp 0: \textnormal{Structure of the test website before any AJAX requests have been executed, regardless of the presence of filter lists.}}
        \label{fig:ajax_before}
    }
    \hfill
    \parbox{0.32\textwidth}{
        % Timestamp 1: all requests execute
        \begin{tblr}{
                width=0.32\textwidth,
                colspec = {|X|X|X|X|X|X|X|X|X|X|X|X|},
                columns = {halign=c},
                column{1-2,4,7-8,10-12} = {wd=1pt},
                column{5-6} = {wd=0pt},
                vline{6} = {0pt},
                vline{1-5,7-13} = {0.5pt},
                colsep=1pt,
                rowsep=0.5pt,
                row{1} = {1pt},
                row{3-4,6-7,9-10} = {2.5em},
                row{2,5,8} = {belowsep=3pt,abovesep=2pt},
                stretch = 0,
                hspan=minimal,
                % 1P AJAX request executes successfully
                cell{3}{1-6,7} = {bg=c1!35},
                cell{6}{2,8} = {bg=c1!35},
                cell{9}{3,9} = {bg=c1!35},
                % 3P AJAX request executes successfully
                cell{4}{1-6,7} = {bg=c2!35},
                cell{7}{2,8} = {bg=c2!35},
                cell{10}{3,9} = {bg=c2!35},
            }
            \SetHline{1-12}{0.5pt}
            \SetCell[c=12]{c} & 2 & 3 & 4 & 5 & 6 & 7 & 8 & 9 & 10 & 11 & 12 \\
            \SetHline{7-11}{0.5pt}
            \SetCell[c=5]{c} \textbf{First-Party Body} & 2 & 3 & 4 & 5 & & \SetCell[c=5]{c} \textbf{Third-Party iframe} & 2 & 3 & 4 & 5 & \\
            \SetCell[c=5]{c} {\texttt{firstparty.com} \\[-1pt] AJAX executed} & 2 & 3 & 4 & 5 & & \SetCell[c=5]{c} {\texttt{firstparty.com} \\[-1pt] AJAX executed} & 2 & 3 & 4 & 5 & \\
            \SetCell[c=5]{c} {\texttt{thirdparty.com} \\[-1pt] AJAX executed} & 2 & 3 & 4 & 5 & & \SetCell[c=5]{c} {\texttt{thirdparty.com} \\[-1pt] AJAX executed} & 2 & 3 & 4 & 5 & \\
            \SetHline{2-4,8-10}{0.5pt}
            & \SetCell[c=3]{c} \textbf{First-Party LF} & 3 & 4 & & & & \SetCell[c=3]{c} \textbf{Third-Party LF} & 3 & 4 & & \\
            & \SetCell[c=3]{c} {\texttt{firstparty.com} \\[-1pt] AJAX executed} & 3 & 4 & & & & \SetCell[c=3]{c} {\texttt{firstparty.com} \\[-1pt] AJAX executed} & 3 & 4 & & \\
            & \SetCell[c=3]{c} {\texttt{thirdparty.com} \\[-1pt] AJAX executed} & 3 & 4 & & & & \SetCell[c=3]{c} {\texttt{thirdparty.com} \\[-1pt] AJAX executed} & 3 & 4 & & \\
            \SetHline{3,9}{1pt}
            & & \textbf{First-Party \\[-1pt] Nested LF} & & & & & & \textbf{Third-Party \\[-1pt] Nested LF} & & & \\
            & & {\texttt{firstparty.com} \\[-1pt] AJAX executed} & & & & & & {\texttt{firstparty.com} \\[-1pt] AJAX executed} & & & \\
            & & {\texttt{thirdparty.com} \\[-1pt] AJAX executed} & & & & & & {\texttt{thirdparty.com} \\[-1pt] AJAX executed} & & & \\
            \SetHline{3,9}{1pt}
            & \SetCell[c=3]{c} & 2 & 3 & & & & \SetCell[c=3]{c} & 3 & 4 & & \\
            \SetHline{2-4,8-10}{0.5pt}
            \SetCell[c=5]{c} & 2 & 3 & 4 & 5 & & \SetCell[c=5]{c} & 2 & 3 & 4 & 5 & \\
            \SetHline{7-11}{0.5pt}
            \SetCell[c=12]{c} & 2 & 3 & 4 & 5 & 6 & 7 & 8 & 9 & 10 & 11 & 12 \\
            \SetHline{1-12}{0.5pt}
        \end{tblr}
        \subcaption{Timestamp 1a: \textnormal{Structure of the test website \textit{without} filter list rules after all AJAX requests have been executed.}}
        \label{fig:ajax_after}
    }
    \hfill
    \parbox{0.32\textwidth}{
        % Timestamp 1a: 3P AJAX requests redirected
        \begin{tblr}{
                width=0.32\textwidth,
                colspec = {|X|X|X|X|X|X|X|X|X|X|X|X|},
                columns = {halign=c},
                column{1-2,4,7-8,10-12} = {wd=1pt},
                column{5-6} = {wd=0pt},
                vline{6} = {0pt},
                vline{1-5,7-13} = {0.5pt},
                colsep=1pt,
                rowsep=0.5pt,
                row{1} = {1pt},
                row{3-4,6-7,9-10} = {2.5em},
                row{2,5,8} = {belowsep=3pt,abovesep=2pt},
                stretch = 0,
                hspan=minimal,
                % 1P AJAX request executes successfully
                cell{3}{1-6,7} = {bg=c1!35},
                cell{6}{2,8} = {bg=c1!35},
                cell{9}{3,9} = {bg=c1!35},
            }
            \SetHline{1-12}{0.5pt}
            \SetCell[c=12]{c} & 2 & 3 & 4 & 5 & 6 & 7 & 8 & 9 & 10 & 11 & 12 \\
            \SetHline{7-11}{0.5pt}
            \SetCell[c=5]{c} \textbf{First-Party Body} & 2 & 3 & 4 & 5 & & \SetCell[c=5]{c} \textbf{Third-Party iframe} & 2 & 3 & 4 & 5 & \\
            \SetCell[c=5]{c} {\texttt{firstparty.com} \\[-1pt] AJAX executed} & 2 & 3 & 4 & 5 & & \SetCell[c=5]{c} {\texttt{firstparty.com} \\[-1pt] AJAX executed} & 2 & 3 & 4 & 5 & \\
            \SetCell[c=5]{c} {[noop text]} & 2 & 3 & 4 & 5 & & \SetCell[c=5]{c} {[noop text]} & 2 & 3 & 4 & 5 & \\
            \SetHline{2-4,8-10}{0.5pt}
            & \SetCell[c=3]{c} \textbf{First-Party LF} & 3 & 4 & & & & \SetCell[c=3]{c} \textbf{Third-Party LF} & 3 & 4 & & \\
            & \SetCell[c=3]{c} {\texttt{firstparty.com} \\[-1pt] AJAX executed} & 3 & 4 & & & & \SetCell[c=3]{c} {\texttt{firstparty.com} \\[-1pt] AJAX executed} & 3 & 4 & & \\
            & \SetCell[c=3]{c} {[noop text]} & 3 & 4 & & & & \SetCell[c=3]{c} {[noop text]} & 3 & 4 & & \\
            \SetHline{3,9}{1pt}
            & & \textbf{First-Party \\[-1pt] Nested LF} & & & & & & \textbf{Third-Party \\[-1pt] Nested LF} & & & \\
            & & {\texttt{firstparty.com} \\[-1pt] AJAX executed} & & & & & & {\texttt{firstparty.com} \\[-1pt] AJAX executed} & & & \\
            & & {[noop text]} & & & & & & {[noop text]} & & & \\
            \SetHline{3,9}{1pt}
            & \SetCell[c=3]{c} & 2 & 3 & & & & \SetCell[c=3]{c} & 3 & 4 & & \\
            \SetHline{2-4,8-10}{0.5pt}
            \SetCell[c=5]{c} & 2 & 3 & 4 & 5 & & \SetCell[c=5]{c} & 2 & 3 & 4 & 5 & \\
            \SetHline{7-11}{0.5pt}
            \SetCell[c=12]{c} & 2 & 3 & 4 & 5 & 6 & 7 & 8 & 9 & 10 & 11 & 12 \\
            \SetHline{1-12}{0.5pt}
        \end{tblr}
        \subcaption{Timestamp 1b: \textnormal{Structure of the test website \textit{with} filter list rules redirecting AJAX requests for \texttt{thirdparty.com}.}}
        \label{fig:ajax_3p}
    }
    \caption{Structure of the test website for resource replacement (a) before and (b) after AJAX request execution, and (c) the expected behavior of our test for RQ2. \textnormal{Cells highlighted in yellow indicate successful AJAX requests from \texttt{firstparty.com}, and cells highlighted in red indicate successful AJAX requests from \texttt{thirdparty.com}.}}
\end{table*}

% Revert commands
\normalsize
\renewcommand{\tablename}{Table}

%% file: tables/scriptlet_and_cosmetic.tex
% Commands to revert later
\small

\begin{table*}[t]
    \captionsetup{type=figure}
    \parbox{0.48\textwidth}{
        % Cosmetic: nothing hidden
        \begin{tblr}{
                width=0.48\textwidth,
                colspec = {|X|X|X|X|X|X|X|X|X|X|X|X|},
                columns = {halign=c},
                column{1-2,4,7-8,10-12} = {wd=1pt},
                column{5-6} = {wd=0pt},
                vline{6} = {0pt},
                vline{1-5,7-13} = {0.5pt},
                colsep=1pt,
                rowsep=0.5pt,
                row{1} = {1pt},
                % row{3-4,6-7,9-10} = {2.5em},
                row{2,4,6} = {belowsep=3pt,abovesep=2pt},
                row{3,5,7} = {1em},
                stretch = 0,
                hspan=minimal,
                % Color 1P visible
                cell{3}{1-6} = {bg=c1!35},
                cell{5}{2} = {bg=c1!35},
                cell{7}{3} = {bg=c1!35},
                % Color 3P visible
                cell{3}{7} = {bg=c2!35},
                cell{5}{8} = {bg=c2!35},
                cell{7}{9} = {bg=c2!35},
            }
            \SetHline{1-12}{0.5pt}
            \SetCell[c=12]{c} & 2 & 3 & 4 & 5 & 6 & 7 & 8 & 9 & 10 & 11 & 12 \\
            \SetHline{7-11}{0.5pt}
            \SetCell[c=5]{c} \textbf{First-Party Body} & 2 & 3 & 4 & 5 & & \SetCell[c=5]{c} \textbf{Third-Party iframe} & 2 & 3 & 4 & 5 & \\
            \SetCell[c=5]{c} {Visible text} & 2 & 3 & 4 & 5 & & \SetCell[c=5]{c} {Visible text} & 2 & 3 & 4 & 5 & \\
            \SetHline{2-4,8-10}{0.5pt}
            & \SetCell[c=3]{c} \textbf{First-Party LF} & 3 & 4 & & & & \SetCell[c=3]{c} \textbf{Third-Party LF} & 3 & 4 & & \\
            & \SetCell[c=3]{c} {Visible text} & 3 & 4 & & & & \SetCell[c=3]{c} {Visible text} & 3 & 4 & & \\
            \SetHline{3,9}{1pt}
            & & \textbf{First-Party Nested LF} & & & & & & \textbf{Third-Party Nested LF} & & & \\
            & & {Visible text} & & & & & & {Visible text} & & & \\
            \SetHline{3,9}{1pt}
            & \SetCell[c=3]{c} & 2 & 3 & & & & \SetCell[c=3]{c} & 3 & 4 & & \\
            \SetHline{2-4,8-10}{0.5pt}
            \SetCell[c=5]{c} & 2 & 3 & 4 & 5 & & \SetCell[c=5]{c} & 2 & 3 & 4 & 5 & \\
            \SetHline{7-11}{0.5pt}
            \SetCell[c=12]{c} & 2 & 3 & 4 & 5 & 6 & 7 & 8 & 9 & 10 & 11 & 12 \\
            \SetHline{1-12}{0.5pt}
        \end{tblr}
        \subcaption{No filter list rules applied.}
        \label{fig:cosmetic_unmodified}
    }
    \hfill
    \parbox{0.48\textwidth}{
        % Cosmetic: 3P hidden
        \begin{tblr}{
                width=0.48\textwidth,
                colspec = {|X|X|X|X|X|X|X|X|X|X|X|X|},
                columns = {halign=c},
                column{1-2,4,7-8,10-12} = {wd=1pt},
                column{5-6} = {wd=0pt},
                vline{6} = {0pt},
                vline{1-5,7-13} = {0.5pt},
                colsep=1pt,
                rowsep=0.5pt,
                row{1} = {1pt},
                % row{3-4,6-7,9-10} = {2.5em},
                row{2,4,6} = {belowsep=3pt,abovesep=2pt},
                row{3,5,7} = {1em},
                stretch = 0,
                hspan=minimal,
                % Color 1P visible
                cell{3}{1-6} = {bg=c1!35},
                cell{5}{2} = {bg=c1!35},
                cell{7}{3} = {bg=c1!35},
            }
            \SetHline{1-12}{0.5pt}
            \SetCell[c=12]{c} & 2 & 3 & 4 & 5 & 6 & 7 & 8 & 9 & 10 & 11 & 12 \\
            \SetHline{7-11}{0.5pt}
            \SetCell[c=5]{c} \textbf{First-Party Body} & 2 & 3 & 4 & 5 & & \SetCell[c=5]{c} \textbf{Third-Party iframe} & 2 & 3 & 4 & 5 & \\
            \SetCell[c=5]{c} Visible text & 2 & 3 & 4 & 5 & & \SetCell[c=5]{c} [Hidden text] & 2 & 3 & 4 & 5 & \\
            \SetHline{2-4,8-10}{0.5pt}
            & \SetCell[c=3]{c} \textbf{First-Party LF} & 3 & 4 & & & & \SetCell[c=3]{c} \textbf{Third-Party LF} & 3 & 4 & & \\
            & \SetCell[c=3]{c} Visible text & 3 & 4 & & & & \SetCell[c=3]{c} [Hidden text] & 3 & 4 & & \\
            \SetHline{3,9}{1pt}
            & & \textbf{First-Party Nested LF} & & & & & & \textbf{Third-Party Nested LF} & & & \\
            & & Visible text & & & & & & [Hidden text] & & & \\
            \SetHline{3,9}{1pt}
            & \SetCell[c=3]{c} & 2 & 3 & & & & \SetCell[c=3]{c} & 3 & 4 & & \\
            \SetHline{2-4,8-10}{0.5pt}
            \SetCell[c=5]{c} & 2 & 3 & 4 & 5 & & \SetCell[c=5]{c} & 2 & 3 & 4 & 5 & \\
            \SetHline{7-11}{0.5pt}
            \SetCell[c=12]{c} & 2 & 3 & 4 & 5 & 6 & 7 & 8 & 9 & 10 & 11 & 12 \\
            \SetHline{1-12}{0.5pt}
        \end{tblr}
        \subcaption{Expected behavior for hiding elements on \texttt{thirdparty.com}.}
        \label{fig:cosmetic_expected}
    }
    \vskip -1em
    \caption{Representation of (a) our test website for cosmetic filtering and (b) the expected behavior for RQ4.  \textnormal{Cells highlighted in yellow indicate elements created by \texttt{firstparty.com}, and cells highlighted in red indicate elements from \texttt{thirdparty.com}.}}
    \vskip -1em
\end{table*}

% Revert commands
\normalsize

%% file: tables/tool_results.tex
\begin{table*}[h]
\centering
\resizebox{\textwidth}{!}{ 
\begin{tabular}{c c c c c c}
\toprule
\textbf{Tool} & \textbf{Platform} & \begin{tabular}[c]{@{}c@{}}\textbf{Request Blocking}\\(RQ1)\end{tabular} & \begin{tabular}[c]{@{}c@{}}\textbf{Resource Replacement}\\(RQ2)\end{tabular} & \begin{tabular}[c]{@{}c@{}}\textbf{Scriptlet Injection}\\(RQ3)\end{tabular} & \begin{tabular}[c]{@{}c@{}}\textbf{Cosmetic Filters}\\(RQ4)\end{tabular} \\ \midrule
\multirow{3}{*}{AdBlock Plus} & Chrome Extension & \emptycirc & \emptycirc & \halfcirc & \emptycirc \\ 
 & Firefox Extension & \emptycirc & \emptycirc & \emptycirc & \emptycirc \\ 
 & iOS & \emptycirc & N/A & N/A & \fullcirc \\ \midrule
\multirow{3}{*}{uBlock Origin} & Chrome Extension & \emptycirc & \emptycirc & \halfcirc & \emptycirc \\ 
 & Firefox Extension & \emptycirc & \emptycirc & \emptycirc & \emptycirc \\
 & \begin{tabular}[c]{@{}c@{}}Chrome MV3\\(uBlock Origin Lite)\end{tabular} & \emptycirc & \emptycirc & \hspace{4pt}\emptycirc$^*$ & \fullcirc \\ \midrule
\multirow{3}{*}{AdGuard} & Chrome Extension & \emptycirc & \emptycirc & \fullcirc & \fullcirc \\ 
 & Firefox Extension & \emptycirc & \emptycirc & \fullcirc & \fullcirc \\
 & iOS & \emptycirc & N/A & N/A & \fullcirc \\ \midrule
\multirow{3}{*}{Brave Browser} & Desktop & \emptycirc & \emptycirc & \fullcirc & \fullcirc \\
 & iOS & \fullcirc & \fullcirc & \fullcirc & \fullcirc \\
 & Android & \emptycirc & \emptycirc & \fullcirc & \fullcirc \\ \midrule 
\multirow{5}{*}{DuckDuckGo} & Chrome Extension & \emptycirc & \emptycirc & N/A & N/A \\
& Firefox Extension & \emptycirc & \emptycirc & N/A & N/A \\
& Desktop & \hspace{4pt}\fullcirc\textsuperscript{\textdagger} & \emptycirc & N/A & N/A \\ 
& iOS & \emptycirc & \emptycirc & N/A & N/A \\ 
& Android & \emptycirc & \emptycirc & N/A & N/A \\ \midrule
% \begin{tabular}[c]{@{}c@{}}Firefox Enhanced Tracking\\ Protection (Strict Mode)\end{tabular} & MacOS & \emptycirc & N/A & N/A & N/A \\
 \multirow{1}{*}{Safari Content Blocker} & MacOS & \emptycirc & N/A & N/A & \fullcirc \\ 
\bottomrule
\end{tabular}
}
    \vskip 1em
\caption{\textbf{Results showing which tools can be evaded for each capability}. \textnormal{\fullcirc{} indicates the tool is vulnerable for that capability, \emptycirc{} indicates the tool is not vulnerable for that capability and \halfcirc{} indicates that the capability is inconsistently effective. $^*$uBlock Origin Lite was vulnerable before commit 520f81f; the issue was patched during the disclosure process for other uBlock Origin issues. \textsuperscript{\textdagger}We find DuckDuckGo's request blocking cannot be evaded, but websites can evade the privacy harms of their website being reported to users.}}
\vskip -2em
\label{tab:vuln-results}
\end{table*}

%% file: tables/safari.tex
% Commands to revert later
\renewcommand{\tablename}{Figure}
\small

\begin{table*}[t]
    \captionsetup{type=figure}
    \parbox{0.495\textwidth}{
        \begin{tblr}{
                width=0.495\textwidth,
                colspec = {|X|X|X|X|X|X|X|X|X|X|X|X|},
                columns = {halign=c},
                column{1-2,4,7-8,10-12} = {wd=1pt},
                column{5-6} = {wd=0pt},
                vline{6} = {0pt},
                vline{1-5,7-13} = {0.5pt},
                colsep=1pt,
                rowsep=0.5pt,
                rows = {valign=m},
                row{1} = {1pt},
                row{2,5,8} = {belowsep=3pt,abovesep=2pt},
                row{3-4,6-7,9-10} = {1em},
                stretch = 0,
                hspan=minimal,
                % 1P script executes successfully
                cell{3}{1-6,7} = {bg=c1!35},
                cell{6}{2,8} = {bg=c1!35},
                cell{9}{3,9} = {bg=c1!35},
            }
            \SetHline{1-12}{0.5pt}
            \SetCell[c=12]{c} & 2 & 3 & 4 & 5 & 6 & 7 & 8 & 9 & 10 & 11 & 12 \\
            \SetHline{7-11}{0.5pt}
            \SetCell[c=5]{c} \textbf{First-Party Body} & 2 & 3 & 4 & 5 & & \SetCell[c=5]{c} \textbf{Third-Party iframe} & 2 & 3 & 4 & 5 & \\
            \SetCell[c=5]{c} {\texttt{firstparty.com} script executed} & 2 & 3 & 4 & 5 & & \SetCell[c=5]{c} {\texttt{firstparty.com} script executed} & 2 & 3 & 4 & 5 & \\
            \SetCell[c=5]{c} {[no text]} & 2 & 3 & 4 & 5 & & \SetCell[c=5]{c} {[no text]} & 2 & 3 & 4 & 5 & \\
            \SetHline{2-4,8-10}{0.5pt}
            & \SetCell[c=3]{c} \textbf{First-Party LF} & 3 & 4 & & & & \SetCell[c=3]{c} \textbf{Third-Party LF} & 3 & 4 & & \\
            & \SetCell[c=3]{c} {\texttt{firstparty.com} script executed} & 3 & 4 & & & & \SetCell[c=3]{c} {\texttt{firstparty.com} script executed} & 3 & 4 & & \\
            & \SetCell[c=3]{c} {[no text]} & 3 & 4 & & & & \SetCell[c=3]{c} {[no text]} & 3 & 4 & & \\
            \SetHline{3,9}{1pt}
            & & \textbf{First-Party Nested LF} & & & & & & \textbf{Third-Party Nested LF} & & & \\
            & & {\texttt{firstparty.com} script executed} & & & & & & {\texttt{firstparty.com} script executed} & & & \\
            & & {[no text]} & & & & & & {[no text]} & & & \\
            \SetHline{3,9}{1pt}
            & \SetCell[c=3]{c} & 2 & 3 & & & & \SetCell[c=3]{c} & 3 & 4 & & \\
            \SetHline{2-4,8-10}{0.5pt}
            \SetCell[c=5]{c} & 2 & 3 & 4 & 5 & & \SetCell[c=5]{c} & 2 & 3 & 4 & 5 & \\
            \SetHline{7-11}{0.5pt}
            \SetCell[c=12]{c} & 2 & 3 & 4 & 5 & 6 & 7 & 8 & 9 & 10 & 11 & 12 \\
            \SetHline{1-12}{0.5pt}
        \end{tblr}
        \subcaption{Third-party request blocking using Safari Content Blocker.}
        \label{fig:safari_blockreq}
    }
        \hfill
        \parbox{0.495\textwidth}{
        % All resources load
        \begin{tblr}{
                width=0.495\textwidth,
                colspec = {|X|X|X|X|X|X|X|X|X|X|X|X|},
                columns = {halign=c},
                column{1-2,4,7-8,10-12} = {wd=1pt},
                column{5-6} = {wd=0pt},
                vline{6} = {0pt},
                vline{1-5,7-13} = {0.5pt},
                colsep=1pt,
                rowsep=0.5pt,
                rows = {valign=m},
                row{1} = {1pt},
                row{2,5,8} = {belowsep=3pt,abovesep=2pt},
                row{3-4,6-7,9-10} = {1em},
                stretch = 0,
                hspan=minimal,
                % 1P script executes successfully
                cell{3}{1-6} = {bg=c1!35},
                cell{6}{2} = {bg=c1!35},
                cell{9}{3} = {bg=c1!35},
                % 3P script executes successfully
                cell{4}{7} = {bg=c2!35},
                cell{7}{8} = {bg=c2!35},
                cell{10}{9} = {bg=c2!35},
            }
            \SetHline{1-12}{0.5pt}
            \SetCell[c=12]{c} & 2 & 3 & 4 & 5 & 6 & 7 & 8 & 9 & 10 & 11 & 12 \\
            \SetHline{7-11}{0.5pt}
            \SetCell[c=5]{c} \textbf{First-Party Body} & 2 & 3 & 4 & 5 & & \SetCell[c=5]{c} \textbf{Third-Party iframe} & 2 & 3 & 4 & 5 & \\
            \SetCell[c=5]{c} {\texttt{firstparty.com} script executed} & 2 & 3 & 4 & 5 & & \SetCell[c=5]{c} {[no text]} & 2 & 3 & 4 & 5 & \\
            \SetCell[c=5]{c} {[no text]} & 2 & 3 & 4 & 5 & & \SetCell[c=5]{c} {\texttt{thirdparty.com} script executed} & 2 & 3 & 4 & 5 & \\
            \SetHline{2-4,8-10}{0.5pt}
            & \SetCell[c=3]{c} \textbf{First-Party LF} & 3 & 4 & & & & \SetCell[c=3]{c} \textbf{Third-Party LF} & 3 & 4 & & \\
            & \SetCell[c=3]{c} {\texttt{firstparty.com} script executed} & 3 & 4 & & & & \SetCell[c=3]{c} {[no text]} & 3 & 4 & & \\
            & \SetCell[c=3]{c} {[no text]} & 3 & 4 & & & & \SetCell[c=3]{c} {\texttt{thirdparty.com} script executed} & 3 & 4 & & \\
            \SetHline{3,9}{1pt}
            & & \textbf{First-Party Nested LF} & & & & & & \textbf{Third-Party Nested LF} & & & \\
            & & {\texttt{firstparty.com} script executed} & & & & & & {[no text]} & & & \\
            & & {[no text]} & & & & & & {\texttt{thirdparty.com} script executed} & & & \\
            \SetHline{3,9}{1pt}
            & \SetCell[c=3]{c} & 2 & 3 & & & & \SetCell[c=3]{c} & 3 & 4 & & \\
            \SetHline{2-4,8-10}{0.5pt}
            \SetCell[c=5]{c} & 2 & 3 & 4 & 5 & & \SetCell[c=5]{c} & 2 & 3 & 4 & 5 & \\
            \SetHline{7-11}{0.5pt}
            \SetCell[c=12]{c} & 2 & 3 & 4 & 5 & 6 & 7 & 8 & 9 & 10 & 11 & 12 \\
            \SetHline{1-12}{0.5pt}
        \end{tblr}
        \subcaption{Expected behavior for blocking third-party requests.}
        \label{fig:blockreqs_3p}
    }
    % \vskip -1em
    \caption{Structure of our test website for blocking requests. \textnormal{We find (a) Safari Content Blocker's interpretation of third-party requests differs from (b) all other non-vulnerable tools, which matches the expected behavior.}}
    % \vskip -1em
\end{table*}

% Revert commands
\normalsize
\renewcommand{\tablename}{Table}

%% file: tables/disclosure.tex
% \usepackage{booktabs}
\begin{table*}[t]
    \begin{tabular}{@{}lllll@{}}
    \toprule
    \multicolumn{1}{c}{Browser/Tool} & \multicolumn{1}{c}{Issue}                    & \multicolumn{1}{c}{Report Date} & \multicolumn{1}{c}{Fix Date} & \multicolumn{1}{c}{Report URL} \\ \midrule
    Brave & Scriptlets not injected & 8/24/24 & 12/3/24 & \cite{bravepatch} \\
    Brave & Cosmetic filters not applied & 8/24/24 & 12/3/24 & \cite{bravepatch} \\
    Brave & Resource redirection not working on iOS & 8/24/24 & 12/3/24 & \cite{bravepatch} \\
    Brave & Incorrect request blocking on iOS & 8/24/24 & 12/3/24 & \cite{bravepatch} \\
    DuckDuckGo & Accounting of blocked requests & 9/20/24 & 10/16/24 & \cite{ddgpatch} \\
    AdGuard & Origin miscomputation for scriptlets & 8/17/24 & 10/4/24 & \cite{adguardpatch} \\
    AdGuard & Origin miscomputation for cosmetic filters & 8/17/24 & 10/4/24 & \cite{adguardpatch}  \\
    Apple & Cosmetic filters not applied & 8/17/24 & 3/31/25 & N/A \\
    Apple & Third-party definition does not match others & 8/17/24 & N/A & N/A \\
    AdBlock Plus & Scriptlets not injected & 8/17/24 & N/A & N/A  \\
    AdBlock Plus & Cosmetic filters not applied & 8/17/24 & N/A & N/A  \\
    uBlock Origin & Scriptlets applied inconsistently & 8/17/24 & N/A & N/A \\
    uBlock Origin Lite & Cosmetic filters not applied & 8/20/24 & N/A & N/A \\
    \bottomrule
    \end{tabular}
    \vskip 1em
    \caption{Summary of our responsible disclosure.}
    \label{tab:disclosure}
    \vskip -2em
\end{table*}

%% file: content/ethics_disclosure.tex
\section{Ethics and Disclosure}

We disclosed all 19 vulnerabilities found to the six affected parties through their preferred channels (Table~\ref{tab:disclosure}), and the standard 90-day disclosure period has passed. At time of writing, our findings were acknowledged by all organizations (AdBlock Plus, AdGuard, Apple, Brave, DuckDuckGo and uBlock Origin). 
Moreover, Brave, Apple, AdGuard, and DuckDuckGo have issued patches.
%\footnote{See \url{https://github.com/brave/brave-browser/issues/40649} 
%and \url{https://github.com/brave/brave-browser/issues/40703}}, 
% Apple and AdGuard are currently implementing patches. 
%\todo{First footnote link deanonymizes us} 
% In particular, Brave is introducing checks to see if the current frame is a local frame, and if so, find a non-local parent frame.

For AdGuard, we determined that introducing a flag in the extension's manifest file solves some---but not all---of their issues. The \texttt{match\_origin\_as\_fallback} flag sets the origin of a local frame to the frame that created it~\cite{contentscripts}. Adding this flag to AdGuard's manifest addresses the scriptlet injection bypass, but AdGuard is still vulnerable to the cosmetic filtering bypass because they still attribute the local frame's origin incorrectly. This also highlights how content blockers may have multiple code paths for handling inherited origins, which can further increase the complexity of their codebases and make it harder to fix these errors.
Ultimately, AdGuard's patch identifies local frames and uses the URL of the top-level parent of the local frame to determine whether to inject scriptlets and cosmetic filters~\cite{adguardpatch}.

For uBlock Origin, our disclosure process led the maintainers to discover local-frame bypasses with another product. We found that scriptlets are inconsistently injected into local frames for the uBlock Origin Chrome extension.  Upon disclosing this to the maintainer of uBlock Origin, he then tested uBlock Origin Lite (which we had not yet tested). The maintainer found that scriptlets are not injected in local frames and patched this immediately.\footnote{\url{https://github.com/gorhill/uBlock/commit/520f81f}} The patch for scriptlets involved checking if the current document's (non-inherited) origin is null, and if so, finding the first non-null origin of a parent frame.
The maintainer subsequently helped us set up an environment to subject uBlock Origin Lite to our suite of tests. We confirmed that scriptlets could be evaded before the patch, and that the issue was fixed after the patch. We also found that cosmetic filters are not injected into local frames; we then disclosed this to the maintainer, who does not plan to fix this until receiving user complaints, as he believes the patch will incur high performance overhead~\cite{ubo_disclousre}.

%% file: content/discussion.tex
\section{Discussion}
\label{sec:diss}

We discuss implications of our work for existing research and other tools that may present similar vulnerabilities.

% While our study is limited to content blockers' handling of local frames, we suspect similar issues exist both in different Web-based tools and for other ``corner cases'' of the HTML specification.  Our experience suggests that the
% %We believe there are several root causes for why so many content blocking tools mishandle local frames, such as the 
% complexity of the Web platform
% %, vulnerabilities in widely-used APIs,
% combined with limitations of widely-used frameworks leads developers to craft bespoke work-arounds that are likely to introduce bugs and vulnerabilities. 

%APIs that push developers to sidestep the APIs and make mistakes in the process.

\subsection{Web Complexity}

As the Web has evolved to provide more capabilities and rich functionality, its subtleties have also increased. Local frames and their inherited origins are just one example of the unexpected behaviors that Web developers must anticipate. It is difficult for any single person (or organization) to correctly parse every nuance of the standards governing the Web.

% %First, the complexity of the HTML specification makes it more likely for developers to make mistakes. 
% It is well known that developers struggle with the intricacies of HTML; a 2022 analysis found that 68\% of domains contain violations of the HTML 
% %, underscoring the difficulty in parsing and complying with all the guidelines
% specification~\cite{hantke2022html}.  Unfortunately for content-blocker developers, they must ensure that their interventions work across a large set of websites, taking care to have comprehensive blocking while minimizing site breakage---whether or not the website in question properly adheres to the specification.
% %
%We join the chorus calling for automated tooling and formal verification to aid developers of content blockers to avoid %pitfalls such as the ones we identify.
%
%While a website creator can check for bugs on their own website, 

% Even worse, these inevitable vulnerabilities in popular libraries silently cascade.
%limitations with standard tooling  can both introduce errors in the tools that use the API. We found that 
% In our testing, the cosmetic filtering in both AdGuard and Brave iOS apps can be evaded by local frames, but the base vulnerability lies not in those apps but in Safari's Content Blocking functionality.

Furthermore, Web toolkits that are supposed to aid developers in reducing complexity can actually increase code complexity---and introduce avenues for evasion---as developers try to work around the API to achieve their desired functionality. 
Concretely, Safari Content Blocking is limited to 150,000 rules per extension~\cite{safari_filter_limit}.  Developers at both AdGuard and Brave found ways to apply larger sets of rules, but their approaches each have downsides. AdGuard instantiates several Safari Content Blocker extensions (each of which has an separate 150,000-rule limit), leading to a tedious setup process where users must enable each extension individually~\cite{adguard_setup}. More critically, Brave implements an alternative code path in order to block additional requests, and this code path is vulnerable to evasion by local frames.  Hence, while Safari Content Blocking provides a handy primitive, its inability to scale to the requirements of modern content blockers 
%like Brave and AdGuard---the latter of which requires users to enable multiple different content blocking extensions in %order to load all of their desired filter list rules~\cite{safari_filter_limit}. The fact that Brave and AdGuard struggle to load all of their filter lists with Safari Content Blocking points 
% leads to a broader ecosystem issue.
creates some of the same issues it tries to address.

% The existence of Brave's two separate code paths for request blocking indicates a larger, ecosystem-wide issue.
% The second, vulnerable code path for request blocking is a direct response to the limitations of the Safari Content Blocking tool.
% % The non-vulnerable code path for request blocking implements the Safari Content Blocking tool. However, as we previously mentioned, this tool restricts the number of filter list rules to 150,000 rules~\cite{safari_filter_limit}.
% % ; for context, EasyList and EasyPrivacy (as of August 17, 2024) have a combined 123,931 rules---and these are only some of the filter list rules that Brave loads by default. 
% In order to get around these limitations, Brave added complexity (and easily exploitable bugs) to their codebase. While Safari Content Blocking creates a handy primitive for iOS apps, it does not scale to the requirements of modern content blockers like Brave and AdGuard---the latter of which requires users to enable multiple different content blocking extensions in order to load all of their desired filter list rules~\cite{safari_filter_limit}. The fact that Brave and AdGuard struggle to load all of their filter lists with Safari Content Blocking points to a broader ecosystem issue.

\subsection{Impact on Research}

% Similarly, we expect that much existing Web privacy and security research makes many of the same errors as the tools examined in this work. This has a worrying implication for any understanding of the Web based on existing research. Measurements that misunderstand how \LFs{} are handled on the Web will incorrectly measure how many cookies are shared, created, or stored; mis-attribute responsibility for fingerprinting, tracking, and other behaviors; and potentially make other similar errors. Quantifying the impact on our understanding of the Web (based on existing research) is important work, but beyond the scope of this project.

Many areas of Web research (both general measurement studies and those focused on security and privacy topics) often need to accurately distinguish between first and third parties on the Web. For example, any research that involves filter lists, either directly (e.g.,
the maintenance and exception policies of filter lists~\cite{snyder2020filters}),
or indirectly (e.g., as a set of ground-truth labels for broken websites~\cite{smith2022blocked, le2023autofr, nisenoff2023defining})
requires correctly replicating how browsers determine the ``party-ness'' of
local frames.
Even more broadly, correctly determining ``party-ness'' is important for topics like emulating the security and privacy policies of Web browsers, correctly
measuring and attributing behaviors on Web pages (e.g., for browser fingerprinting measurement~\cite{englehardt2016online,iqbal2021fingerprinting}), and understanding  what parties are reading and writing cookies (e.g., \cite{jueckstock2022measuring}). 

To understand the implications of our findings on existing research, we sampled a small number of papers from top security and privacy conferences focusing on filter list rules or ad measurement~\cite{yeung24analyzing,amjad2021trackersift,le2023autofr,zafar2021understanding,dao2021alternative,siby2022webgraph,zafar2023comparative}. Of the seven papers, only five~\cite{yeung24analyzing,dao2021alternative,amjad2021trackersift,le2023autofr,zafar2021understanding} provided code or pointers to the tools they used. Of these five, we did not find any that correctly computed ``party-ness'' or correctly handled local frames, either because 1) the provided code determines a frame's party (or security origin) incorrectly, or 2) the provided code logs information for future analysis, and the logging code does not capture the needed information to correctly determine ``party-ness'' (i.e., the frame's security origin). As one example, we find that the Adscraper tool~\cite{Zeng_adscraper} (and its use in one recent paper~\cite{yeung24analyzing}) incorrectly checks for ``party-ness'' against the top-level document instead of the containing frame. 
%See Appendix~\ref{appendix:other_research} for more details.

We do not claim that our analysis of a small sample of papers is comprehensive or representative, and conducting a comprehensive study would both be far beyond the scope of this work and require
resources not available to us (e.g., source code, data sets, measurement
raw data). Nevertheless, our sampling of existing research suggests that tools used to conduct
many prior Web security and privacy research studies may contain non-trivial
bugs leading to incorrect results, stemming from understandable---though important---misunderstandings in subtle aspects of browser security and privacy policies.

\subsection{Other Possible Vulnerabilities}

%Beyond local frames, there are many other Web features that could be mishandled by privacy tools. In Section~\ref{sec:bkgd:origins} we listed other iFrame URIs such as the ``javascript'', ``blob'', ``file'', and ``data'' prefixes. 

Beyond content blockers, there are many other classes of privacy-focused Web tools that could similarly mishandle local frames.  We discuss two particularly sensitive ones below.

%, such as Firefox's opt-in blocking of fingerprinting, Safari's blocking of fingerprinting and trackers in Private Browsing mode, and content blocking extensions in other browsers (e.g. Opera). Similarly, Chrome and AdBlock Plus allowlist ads that follow the Better Ads and Acceptable Ads standards, respectively~\cite{chrome_better_ads,acceptable_ads}; local frames may result in such ads being blocked.
%We are heartened that one particularly sensitive class of tools---password managers---seems largely immune. 

\subsubsection{Password managers}
\input{appendix/password_autofill}

\subsubsection{Anti-fingerprinting tools}
We also test whether browser extensions that spoof canvas fingerprints can be evaded by local frames.  Our tests of three extensions considered in a recent study by Nguyen and Vadrevu~\cite{nguyen} find that, while all successfully spoof fingerprints in local frames, one extension fails to achieve the stronger goal of creating consistent per-domain fingerprints. We test Canvas Fingerprint Defender (Chrome extension version 0.2.2), CanvasBlocker (Firefox extension version 1.11), and Canvas Blocker - Fingerprint Protect (Chrome and Firefox extensions version 0.2.1). The latter extension fails to create consistent per-domain fingerprints on Firefox because their method for copying fingerprints from parent frames fails to account for Firefox's extension sandboxing model; we reported this bug to the developer along with a patch. (The Chrome extension successfully creates consistent fingerprints.)

\subsection{Potential Breakage}

\input{content/breakage}

%% file: appendix/password_autofill.tex
%\section{Password Autofill}
\label{appendix:password}

% While evading privacy tools is concerning, our findings suggest similar evasion might be possible for other types of tools as well.  We are heartened that one particularly sensitive class of tool seems largely immune. 
We investigate the possibility that local frames could be exploited to exfiltrate user credentials by tricking password autofill tools into submitting credentials for a different website into a login form in a local frame. In particular, we consider a third-party local frame containing a login form, and check if autofill tools fill in the user's credentials for the first-party website.

Fortunately, in our tests of five browsers' native autofill features and two Chrome extensions, none are vulnerable. Several browsers (Chrome, Brave, Firefox, and Safari) and both Chrome extensions (1Password and LastPass) do not support autofilling credentials into any iframes---obviating concerns about local frames or any other type of iframe.  The DuckDuckGo browser, on the other hand, supports autofilling credentials in iframes and correctly determines the origin of the local frame.  
% While it appears to correctly determine the origin of local frames in this instance, our larger study indicates the DuckDuckGo developers made an error in another code path; the correct handling of local frames remains precarious.
% Brave and Safari don't autofill in iframes
% Firefox doesn't autofill in any iframes. In fact, they don't even show the login forms in the iframe
% DuckDuckGo allows autofill in iframes (on HTTPS connections), but they have correct origin attribution even for nested local frames; no vulnerability
(We note that 1Password's debugger correctly identifies the origin of the local frame as a third-party origin, suggesting that even if it were to autofill in iframes, it would not be vulnerable to evasion. 
We are unable to access similar debugging information for the other password managers.)

%% file: content/breakage.tex
There is always the potential that modifying website behavior---either within or outside of local frames---can cause undesirable user-visible impacts, referred to as ``breakage''. When acknowledging our findings, every content-blocker organization indicated that these were unintentional vulnerabilities, and not choices meant to avoid breaking websites.  In general, when privacy tools cause website breakage, authors add custom-tailored exceptions to their filter lists to restore website functionality---which tools must properly handle within local frames as well.
Hence, we consider whether proper local-frame handling may induce additional breakage.

We sample 50 of the \CrawlRequestsSitesLocalFrameBlocked{} websites that make requests inside of local frames that should be blocked (i.e., the set of websites that may suffer additional breakage when implementing privacy protections within local frames).
We employ a methodology to identify breakage proposed by prior work~\cite{iqbal2021fingerprinting,amjad2021trackersift,snyder2017most}, asking two non-authors to interact with two Chrome browsers, one with the uBlock Origin Lite extension (which correctly implements request blocking, resource replacement, and scriptlet injection in local frames) and one without.
%\footnote{We were unable to build Brave iOS before and after it was patched to test whether its request blocking in local frames would cause breakage; so instead we test whether any sort of request blocking will cause breakage and manually investigate whether local frames are involved.}
The testers interact with each of the 50 websites in both browsers looking for obvious visual differences and the following types of breakage:
%Major breakage includes 
non-functional search bar, menu, page navigation,
%present after the patch and not before the patch; minor breakage includes 
comment sections, reviews, social media widgets, or icons. (Suppressed ads are not considered to be breakage.) 
% I did not have time to do this, but the testers checked for visual differences
% Additionally, we compare screenshots of the webpage 
%and check if the hash of the screenshots are the same; if not, we manually review them 
% to identify any visual differences, aside from missing/different ads.  
Neither tester found any instances of breakage involving local frames.

%We have two non-authors review breakage and their inter-coder reliability (Cohen's kappa) is 0.63. By taking an average of their results, we find that
Each tester found exactly one website that exhibited breakage, but the breakage was inconsistent, i.e., the website worked fine for the other tester---and was not related to local frames.
%1 website has major breakage, 0 websites have minor breakage, and 49 websites do not have any breakage. The reviewers find different websites had breakage; 
One tester observed that \texttt{huffpost.com} failed to load videos embedded within articles when uBlock Origin Lite is installed, while the other reviewer found that the login page of \texttt{modbee.com} occasionally (but not always) caused the browser with uBlock Origin Lite installed to crash.   The video requests on \texttt{huffpost.com} are made within the top-level frame, so any breakage is not related to local-frame handling.  Similarly, while we cannot replicate the \texttt{modbee.com} crash, we confirm that the login page does not create any local frames.
%Upon further investigation, we find that the former case does not use local frames for loading videos (all requests to the video provider are made by the top-level frame) and the latter page does not use any local frames, so we believe these instances of breakage are not due to blocking in local frames.

%% file: content/conclusion.tex
\section{Conclusion}

Content-blocking tools aim to improve users' browsing experience and protect user privacy by blocking trackers and hiding ads, but 
% these actions pose a threat to the revenue of website publishers. We find that publishers at-large are evading popular content blockers through the use of local frames. 
we find that they can be easily evaded.
Our work shows that many popular content blockers confuse the origin of local frames, and therefore do not correctly apply their filter-list rules to local frames. We find 19 vulnerabilities in the Brave Browser, AdBlock Plus, AdGuard, uBlock Origin Lite, DuckDuckGo, and Safari Content Blocking (a primitive used by many privacy-enhancing iOS apps). We also find that these vulnerabilities are being exploited by website publishers to evade content blockers (though we do not know if this evasion is intentional). Local frames are prevalent on more than half of popular websites and \CrawlRequestsSitesBlockedOfSites{} of these popular websites make requests to resources that should be blocked according to popular filter lists. Based on our work, Brave, Safari, DuckDuckGo, and AdGuard have patched their systems.

%% file: content/acks.tex
\section*{Acknowledgments}

We thank Stefan Savage, Miro Haller, Ali Ukani, Paul Chung, Anirudh Canumalla, Cindy Moore, and our anonymous reviewers.
This material is based upon work supported by the National Science Foundation
Graduate Research Fellowship Program under Grant No. DGE-2038238. Any opinions,
findings, and conclusions or recommendations expressed in this material are
those of the author(s) and do not necessarily reflect the views of the National
Science Foundation.

%% file: appendix/fp_apis.tex
\section{Fingerprinting APIs}\label{appendix:fp_apis}

We classify a fingerprinting-related API call as any access to one of the following APIs:

\begin{itemize}
    \item CanvasRenderingContext2D.measureText
    \item HTMLCanvasElement.toDataURL
    \item MediaDevices.enumerateDevices
    \item Navigator: appCodeName.get, appName.get, appVersion.get, bluetooth.get, brave.get, deviceMemory.get, doNotTrack.get, getBattery, globalPrivacyControl.get, hardwareConcurrency.get, language.get, languages.get, maxTouchPoints.get, mediaCapabilities.get, mediaDevices.get, plugins.get, productSub.get, usb.get, userAgent.get, userAgentData.get, vendor.get, vendorSub.get
    \item Screen: availHeight.get, availLeft.get, availTop.get, availWidth.get, colorDepth.get, height.get, isExtended.get, pixelDepth.get, width.get
    \item WebGL2RenderingContext: getExtension, getParameter
    \item WebGLRenderingContext: getExtension, getParameter, getShaderPrecisionFormat
\end{itemize}

%% file: appendix/measurement.tex
\input{tables/third-party-frames-hostnames}

%\section{Additional Measurement Results}
\section{Third-Party Entities}
\label{appendix:measurement}

\input{tables/privacy-suspect-requests-receivers}

Table~\ref{tab:entities_and_hostnames} details the corresponding eTLD+1s for the third-party entities presented in Table~\ref{tab:third_parties}. We report the top-10 entities (URLs are mapped to owning organization using the Disconnect entity list~\cite{disconnect}) that are targeted by these privacy-suspect requests in Table~\ref{tab:requests_receive}, all of which are advertising and analytics companies. We find that Google is the most common entity, contacted by almost 5$\times$ more sites than the next-most-popular entity.

%The focus of our paper is on how local frames are mishandled by privacy tools, and the concrete privacy harms that result from this mishandling. However, in this section we provide a brief analysis of some additional aspects of our measurement study.

% \section{Local Frame Prevalence}% by Website Popularity}
% \label{appendix:measurement}

%\subsection{Third-Party Local Frames}

%\input{tables/third_party_frames}

%We find 14,646 third-party local frames across 2,497 websites crawled.
% On average, these sites contained 5.87 third-party \LFs{}, with the maximum being for a ``Twitter video downloader'' site with 53 third-party \LFs{}.
%We map the URL of the third-party \LF{} to its owning organization using the Disconnect entity list~\cite{disconnect}; if there is no entity for a given URL, we use the URL as the entity. 
%We find 165 unique entities, and report the top 10 that create third-party local frames by the number of sites in Table~\ref{tab:third_parties}. The majority of these companies are advertising and analytics companies.

%% file: tables/third-party-frames-hostnames.tex
\begin{table}[ht]
\small
    \begin{tabular}{ll}
    \toprule
    \multicolumn{2}{c}{Ranks [1--15K)} \\
    \multicolumn{1}{c}{Entity} & \multicolumn{1}{c}{eTLD+1s} \\
    \midrule
    Google & \begin{tabular}[l]{@{}l@{}}[2mdn.net, doubleclick.net, google.com, \\ googlesyndication.com, recaptcha.net, \\ youtube-nocookie.com, youtube.com]\end{tabular} \\
    PubMatic & [pubmatic.com] \\
    Unity & [yellowblue.io] \\
    Cloudflare & [cloudflare.com, cloudflarestream.com] \\
    Amazon & [amazon-adsystem.com, twitch.tv] \\
    Vidoomy & [vidoomy.com] \\
    Datadome & [captcha-delivery.com] \\
    NextMillennium & [nextmillmedia.com] \\
    ConnectAdRealtime & [connectad.io] \\
    Piano & [piano.io, tinypass.com] \\
    \midrule
    
    \multicolumn{2}{c}{Ranks [15K--100K)} \\
    % \multicolumn{1}{c}{Entity} & \multicolumn{1}{c}{eTLD+1's} \\
    \midrule
    Google & \begin{tabular}[l]{@{}l@{}}[2mdn.net, doubleclick.net, google.com, \\ googlesyndication.com, recaptcha.net, \\ youtube-nocookie.com, youtube.com]\end{tabular} \\
    adtrafficquality.google & [adtrafficquality.google] \\
    PubMatic & [pubmatic.com] \\
    Cloudflare & [cloudflare.com] \\
    SeedTag & [seedtag.com] \\
    AdYouLike & [omnitagjs.com] \\
    admatic.de & [admatic.de] \\
    Amadeus & [travelaudience.com] \\
    Amazon & [amazon-adsystem.com] \\
    ConnectAdRealtime & [connectad.io] \\
    
    \midrule
    \multicolumn{2}{c}{Ranks [100K--1M)} \\
    % \multicolumn{1}{c}{Entity} & \multicolumn{1}{c}{eTLD+1's} \\
    \midrule
    Google & \begin{tabular}[l]{@{}l@{}}[2mdn.net, doubleclick.net, google.com, \\ googlesyndication.com, recaptcha.net, \\ youtube-nocookie.com, youtube.com]\end{tabular} \\
    adtrafficquality.google & [adtrafficquality.google] \\
    Cloudflare & [cloudflare.com] \\
    PubMatic & [pubmatic.com] \\
    Amadeus & [travelaudience.com] \\
    SeedTag & [seedtag.com] \\
    Jivox & [jivox.com] \\
    Yandex & [yandex.ru] \\
    Chaturbate & [chaturbate.com] \\
    AdYouLike & [omnitagjs.com] \\
    \bottomrule
    \end{tabular}
    \vskip 1em
\caption{The eTLD+1s for the content loaded into third-party local frames by the top-10 entities as reported in Table~\ref{tab:third_parties}.}
\label{tab:entities_and_hostnames}
\vskip -2em
\end{table}

%% file: tables/privacy-suspect-requests-receivers.tex
\begin{table}[t]
    \centering
    \begin{tabular}{lrr}
    \toprule
    Entity & \# Sites & \# Requests \\
    \midrule
    Google & 1514 & 23644 \\
    Microsoft & 350 & 1283 \\
    PubMatic & 332 & 718 \\
    Integral Ad Science & 332 & 1497 \\
    Criteo & 329 & 1284 \\
    Magnite & 292 & 602 \\
    Taboola & 250 & 498 \\
    IndexExchange & 233 & 349 \\
    LiveIntent & 233 & 373 \\
    Nexxen & 229 & 539 \\
    \bottomrule\\
    \end{tabular}
    \caption{The top-10 entities that receive privacy-suspect requests, i.e. requests from local frames that should be blocked.}
    \label{tab:requests_receive}
\end{table}

%% file: appendix/testing_details.tex
\input{tables/block_requests_1p}

\section{Additional Testing Details}\label{appendix:testdetails}

%\begin{lstlisting}[language=HTML, label=lst:nested, float, floatplacement=h!, deletekeywords={for, FRAME}, caption=HTML code at %\texttt{https://firstparty.com} which creates several iFrames: two local frames (one nested inside the other) and a third-party iFrame that embeds its %own set of nested local frames.]
%<body>
%  <iframe src="about:blank">
%    <!-- Local frame for firstparty.com        Origin should be https://firstparty.com -->
%    <iframe src="about:blank">
%      <!-- Nested local frame for firstparty.com     Origin should be https://firstparty.com -->
%    </iframe>
%  </iframe>
%
%  <iframe src="https://thirdparty.com">
%    <iframe src="about:blank">
%      <!-- Local frame for thirdparty.com            Origin should be https://thirdparty.com -->
%      <iframe src="about:blank">
%        <!-- Nested local frame for thirdparty.com   Origin should be https://thirdparty.com -->
%      </iframe>
%    </iframe>
%  </iframe>
%</body>
%\end{lstlisting}

This section provides additional details that we consider in testing, but that do not affect the results of vulnerable tools.
%
%In each of the four tests described in Section~\ref{sec:test_design}, we add an additional local frame nested inside each existing local frame; this produces a page structure shown in Listing~\ref{lst:nested}. We add 
The nested local frames shown in Listing~\ref{lst:teststructure} test for the case wherein a tool checks for local frames by only considering the local frame's direct parent (not the local frame's non-local ancestor). If this were true, then a tool could be evaded by nested local frames, but not by regular local frames. We do not find that nested local frames can evade the capabilities outlined in Section~\ref{sec:capabilities}. However, we find that nested local frames can trick DuckDuckGo into misreporting the privacy harms of a website (Section~\ref{sec:results:ddg}).

%\subsection{Request Blocking Variants}
\label{appendix:blockreq}

Concretely, in the main body of this work, we only present results for universally blocking requests. However, as noted in Section~\ref{sec:capabilities:req}, content blockers sometimes block requests only when they are loaded in a third-party context. We test two variants of \textbf{RQ1} that check the context in which request are made. First, we consider \textbf{RQ1a}: If you block third-party requests to a resource, does the resource only load in a first-party context? This means the first-party frames should allow the script from \texttt{firstparty.com} and block the script from \texttt{thirdparty.com}. The third-party frames should allow the script from \texttt{thirdparty.com}---since the script is local with respect to the origin of the iframe---and block the script from \texttt{firstparty.com}.  Expected behavior is shown in Figure~\ref{fig:blockreqs_3p}.

Second, we consider \textbf{RQ1b}: If you block first-party requests to a resource, does the resource only load in a third-party context? As shown in Figure~\ref{fig:blockreqs_1p}, the first-party frames will only allow the scripts from \texttt{thirdparty.com}, and the third-party frames will only allow the scripts from \texttt{firstparty.com}.

We do not find any tools that are vulnerable to RQ1a or RQ1b, but not RQ1. However, we find that Safari computes the ``party-ness'' of requests differently than all other tools we study (Section~\ref{sec:results:safari}).

%% file: tables/block_requests_1p.tex
% Commands to revert later
\small

\begin{table}[t]
    \small
    \captionsetup{type=figure}
    \begin{tblr}{
            width=\columnwidth,
            colspec = {|X|X|X|X|X|X|X|X|X|X|X|X|},
            columns = {halign=c},
            column{1-2,4,7-8,10-12} = {wd=1pt},
            column{5-6} = {wd=0pt},
            vline{6} = {0pt},
            vline{1-5,7-13} = {0.5pt},
            colsep=1pt,
            rowsep=0.5pt,
            rows = {valign=m},
            row{1} = {1pt},
            row{2,5,8} = {belowsep=1pt,abovesep=1pt},
            row{3-4,6-7,9-10} = {1em},
            stretch = 0,
            hspan=minimal,
            % 1P script executes successfully
            cell{3}{7} = {bg=c1!35},
            cell{6}{8} = {bg=c1!35},
            cell{9}{9} = {bg=c1!35},
            % 3P script executes successfully
            cell{4}{1-6} = {bg=c2!35},
            cell{7}{2} = {bg=c2!35},
            cell{10}{3} = {bg=c2!35},
        }
        \SetHline{1-12}{0.5pt}
        \SetCell[c=12]{c} & 2 & 3 & 4 & 5 & 6 & 7 & 8 & 9 & 10 & 11 & 12 \\
        \SetHline{7-11}{0.5pt}
        \SetCell[c=5]{c} \textbf{First-Party Body} & 2 & 3 & 4 & 5 & & \SetCell[c=5]{c} \textbf{Third-Party iframe} & 2 & 3 & 4 & 5 & \\
        \SetCell[c=5]{c} {[no text]} & 2 & 3 & 4 & 5 & & \SetCell[c=5]{c} {\texttt{firstparty.com} \\ script executed} & 2 & 3 & 4 & 5 & \\
        \SetCell[c=5]{c} {\texttt{thirdparty.com} \\ script executed} & 2 & 3 & 4 & 5 & & \SetCell[c=5]{c} {[no text]} & 2 & 3 & 4 & 5 & \\
        \SetHline{2-4,8-10}{0.5pt}
        & \SetCell[c=3]{c} \textbf{First-Party LF} & 3 & 4 & & & & \SetCell[c=3]{c} \textbf{Third-Party LF} & 3 & 4 & & \\
        & \SetCell[c=3]{c} {[no text]} & 3 & 4 & & & & \SetCell[c=3]{c} {\texttt{firstparty.com} \\ script executed} & 3 & 4 & & \\
        & \SetCell[c=3]{c} {\texttt{thirdparty.com} \\ script executed} & 3 & 4 & & & & \SetCell[c=3]{c} {[no text]} & 3 & 4 & & \\
        \SetHline{3,9}{1pt}
        & & \textbf{First-Party Nested LF} & & & & & & \textbf{Third-Party Nested LF} & & & \\
        & & {[no text]} & & & & & & {\texttt{firstparty.com} \\ script executed} & & & \\
        & & {\texttt{thirdparty.com} \\ script executed} & & & & & & {[no text]} & & & \\
        \SetHline{3,9}{1pt}
        & \SetCell[c=3]{c} & 2 & 3 & & & & \SetCell[c=3]{c} & 3 & 4 & & \\
        \SetHline{2-4,8-10}{0.5pt}
        \SetCell[c=5]{c} & 2 & 3 & 4 & 5 & & \SetCell[c=5]{c} & 2 & 3 & 4 & 5 & \\
        \SetHline{7-11}{0.5pt}
        \SetCell[c=12]{c} & 2 & 3 & 4 & 5 & 6 & 7 & 8 & 9 & 10 & 11 & 12 \\
        \SetHline{1-12}{0.5pt}
    \end{tblr}
    
    \caption{Expected behavior for blocking first-party requests.}
    \label{fig:blockreqs_1p}
    \vskip 1em
\end{table}

% Revert commands
\normalsize